\documentclass[%
 reprint,
amsmath,amssymb,
aps,
pre,
floatfix,
]{revtex4-2}
\usepackage{caption}
\usepackage{subcaption}
\usepackage{epsfig}
\usepackage[utf8]{inputenc}

\usepackage{xcolor}  

\usepackage{placeins}

\usepackage{graphicx}
\usepackage{dcolumn}
\usepackage{bm}
\usepackage{comment}
\usepackage{multirow}
\usepackage{amsmath}

\usepackage{hyperref}  

\begin{document}

\preprint{APS/123-QED}

\title{Topological Feature Extraction of Scanty Time Series Data: A Data-Driven Approach for Dynamic State Change Detection}

\author{B. Rishab Antosh}
\email{rishab.antosh2019@vitstudent.ac.in}

 \author{Sanjit Das}\thanks{Corresponding author: sanjit.das@vit.ac.in}
\affiliation{%
Division of Physics,
School of Advanced Sciences,
Vellore Institute of Technology, Chennai Campus,
Chennai, Tamil Nadu – 600 127, India
}%

 \author{%
 N. Nirmal Thyagu}\thanks{Corresponding author: nirmalthyagu@mcc.edu.in}
\affiliation{%
Department of Physics, 
Madras Christian College (Autonomous) [Affiliated to the University of Madras], 
Chennai, Tamil Nadu – 600 059, India
 \\
}%

%


\date{\today}

\begin{abstract}
Complex dynamical systems often exhibit phase transitions from periodic to chaotic behavior when the bifurcation parameters change, and timely detection of these critical changes is essential for managing and understanding the dynamics of the system. However, traditional methodologies for estimating the maximal Lyapunov exponent (MLE) have proven to be meaningful only when the systems' governing equations are known or given only the time series, it should be sufficiently long and uniformly sampled. For the latter, the real problem arises when the given time series is scanty or laden with missing data points, where faithful reconstruction of the phase space becomes a challenge. This study tackles the challenge by introducing a novel methodology that leverages tools from Topological Data Analysis (TDA), specifically 0-D sublevel persistence, in conjunction with binary classifiers from Machine Learning (ML) to effectively distinguish the periodic and the chaotic regimes. Sublevel persistence is capable of extracting meaningful topological features from the time series by analyzing the changing minima and maxima. We witness that the topological features obtained from the periodic time series leave a repeating signature, while the features from the chaotic time series appear more scattered. We use three ML classifiers—logistic regression, support vector machines, and K-nearest neighbors—to identify the patterns in the topological features that distinguish the periodic and chaotic regimes. The ML classifier models are trained on these topological features and are evaluated using the confusion matrix and performance metric scores. We validated the selection of hyperparameter values by employing the K-fold cross-validation technique, yielding high average accuracy scores exceeding $90\%$ for all the classifiers. We use our trained ML classifiers as binary score quantifiers to classify new time series data, where a score of 0 indicates that it is a periodic time series and a score of 1 indicates that it is a chaotic time series. First, we assess the effectiveness of our methodology on well-studied systems: the Duffing, Rossler, and Lorenz systems. The classification results exhibit clear distinctions between periodic and chaotic regimes, with transitions closely aligning with the MLE, thereby ascertaining the reliability of our approach. Further, we employ our methodology on real-world ECG data to demonstrate its ability to classify normal and abnormal heartbeats. Furthermore, we evaluated the results of this classification using statistical metrics that show promise and potential for further improvement in future studies. We believe that our methodology will be particularly helpful to experimentalists dealing with scanty or sparse time series data.

 
\end{abstract}

  
\maketitle


\section{Introduction}\
\label{sec:intro}
The hallmark of complex dynamical systems is their tendency to undergo bifurcations as system parameters are varied. Identifying these critical transitions, especially marking the onset of chaotic behavior, continues to be a challenging problem in the field of nonlinear dynamics. In certain cases \cite{khasawneh2016chatter,sujith2020complex,garfinkel1997quasiperiodicity} , these transitions towards chaos lead to unpredictability and undesirable performance of the systems, marking the importance of detecting their transitions. Consequently, the detection of chaotic behavior in dynamical systems has been extensively studied in the field. 

Traditionally, the maximal Lyapunov exponent (MLE) \cite{abarbanel1993analysis, eckmann1986liapunov} has been the marquee method for identifying chaos, which is determined by finding the rate of separation of infinitesimally close trajectories if the underlying equations are known. Alternatively, if the governing equations are not known, the MLE \cite{wolf1985determining, sano1985measurement} can still be determined by reconstructing the phase space of the attractor directly from the time series data using Takens's embedding theorem \cite{takens2006detecting}. Alternative techniques include the work of Weibe et al. \cite{wiebe2012heuristic}, which makes use of the discrete Fourier transform (DFT) to analyze the frequency spectrum of the system's time series data. This characterizes the system by counting the number of peaks, where the chaotic time series produces a large number of peaks while periodic ones produce fewer peaks separated by a distance. Another commonly used method is the 0-1 test for chaos \cite{gottwald2004new, gottwald2009implementation} which transforms the time series into two auxiliary variables—the p-q plane that captures the system's dynamics—where 0 corresponds to the periodic while 1 corresponds to the chaotic regime. More recently, tools from topological data analysis (TDA), such as persistent homology (PH), have shown promise in characterizing transitions in complex systems. For instance, the work of Mittal et al. \cite{mittal2017topological} applied persistent homology (PH) on reconstructed phase space to detect a bifurcation route to chaos. 

Although the above-mentioned methodologies have been successful in characterizing the system dynamics using a single state variable (time series), they exhibit intrinsic limitations when the time series data are limited and scanty. Most existing methods \cite{mittal2017topological,antosh2024characterization,guzel2022detecting,khasawneh2018chatter} either inherently take in regularly sampled time series or rely on state space reconstruction techniques, which themselves pose a challenge when time series data is prone to sampling irregularities. Dealing with scanty time series data is important because real-world experimental data are often prone to missing elements or irregularities in sampling due to varying experimental conditions, equipment failure, or human error. Subsequently, the literature states that the amount of data lost correlates with the ability to characterize the system \cite{pavlova2019effects}. However, one can argue that the interpolation and resampling techniques can be employed to obtain evenly sampled data, but computing and validating these techniques can be tedious and time-consuming. Hence, there arises a need for a methodology that can demarcate or characterize chaotic regimes directly from sparse or scanty time series data. To address this gap, Kulp \cite{kulp2013detecting} devised a modified version of the heuristic DFT method \cite{wiebe2012heuristic}. Two modifications have been introduced: the use of the Lomb-Scargle Periodogram (LSP) instead of the DFT and the implementation of a point cloud plot. The point cloud plot reveals a descending staircase pattern for periodic orbits and a decreasing exponential pattern for chaotic orbits. Despite its advantages, determining the cut-off threshold for accurate results remains unclear. 

In our work, we leverage tools from TDA and machine learning (ML) to analyze and classify the scanty time series data. $0-D$ Sub-level set homology \cite{munkres2018elements, munch2017user, ghrist2014elementary}, a tool from TDA, is known to capture the changes in the topology of a real-valued function. We make use of this tool to extract the topological features from the persistence diagrams for the periodic and chaotic scanty time series and analyze the patterns they exhibit. 

Upon investigation, we notice that the extracted features from the periodic time series data exhibit overlapping, repeating patterns, whereas those from the chaotic time series display irregular, scattered patterns. Now, it boils down to a pattern recognition problem, and it necessitates the use of an advanced tool that is capable of learning and differentiating between these distinct pattern structures. Hence, this prompts us to use ML classifiers, specifically binary classifiers in our case, to distinguish the patterns that these topological features leave. However, persistence diagrams do not have a fixed size, and this leads to variable-length topological features (for different time series). Hence, this prompts us to convert the persistence diagrams into vectorized representations, which in turn can be sent to the classifiers to learn and predict them into the respective labels. We opt for three classifiers: LR, SVM, and kNN, to compare and contrast, thereby helping us to pick the best classifier to do the job. These trained classifiers act as binary score quantifiers that can distinguish new scanty time series by assigning 0 if they are periodic and 1 if they are chaotic.

The overview of this paper is as follows: In section \ref{sec:TDA}, we provide the introduction of TDA and 0-D sublevel set persistence. We explain the filtration process using synthetic time series data and show how changing minima and maxima are captured as topological features. We extend the same analysis to the time series obtained from dynamical systems and show how the topological features obtained from the periodic and chaotic time series leave a pattern to distinguish them. In section \ref{sec:ML}, we introduce three binary classifiers of logistic regression (LR), K-nearest neighbors (kNN), and support vector machines (SVM), and their evaluation metrics to assess the models' performance, thereby helping us to compare them. Additionally, we explain the significance of K-fold cross-validation in ascertaining the hyperparameter choices made. We make use of trained classifiers as a binary score quantifier to classify the new unseen time series data. In section \ref{sec:Results}, we apply our proposed framework to well-known systems: the Duffing, Rossler, and Lorenz systems, followed by the real-world ECG data, where we compare our classifiers' performance using the metrics, and we benchmark the ML-classified time series with the standard MLE frameworks to validate our classification. Lastly, in section \ref{sec:Conc} we conclude our study and provide the implications of our study. Also, we provide the future direction of the study for the researchers to work on.

\begin{figure*}
         \centering
         \includegraphics[width=1\textwidth]{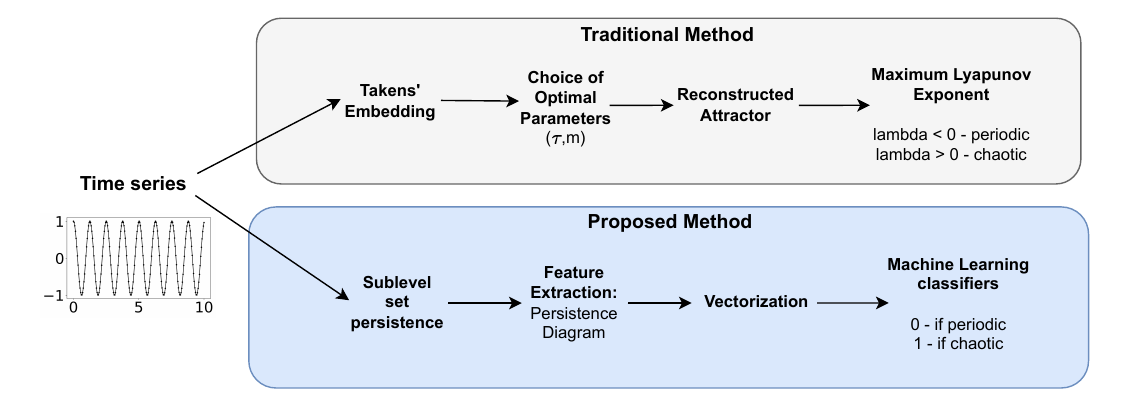}
         \caption{Graphical abstract}
         \label{fig:graphical_abs}
\end{figure*}

\section{Topological Data Analysis} 
\label{sec:TDA}
Topological Data Analysis (TDA), a relatively new field of research, has provided breakthroughs in many areas, including medicine \cite{nicolau2011topology}, robotics \cite{bhattacharya2015persistent}, and dynamical systems \cite{mittal2017topological}. TDA aims to analyze the shape and topology of the data under study by analyzing properties like connectedness, loops, and voids. TDA combines algebraic topology, statistics, data analysis, and computational topology to provide a meaningful qualitative analysis of the dataset under study. Persistent Homology (PH), \cite{edelsbrunner2008persistent, zomorodian2001computing} a powerful tool in the branch of TDA, is widely used in many places, and it is mainly known for its robust nature, stability to noise, and easy interpretability of topological features obtained. Our region of interest lies in the domain of time series data analysis. Within this context, sublevel set persistence \cite{munkres2018elements, munch2017user, ghrist2014elementary} from TDA has shown promise in extracting meaningful information from one-dimensional data. Researchers have made use of sublevel set persistence in different scientific disciplines. For instance, it has been employed to quantify the architectural features of prostate cancer \cite{lawson2019persistent}, to analyze the topology of the energy landscape for n-alkanes \cite{mirth2021representations},  and to demarcate the noise from the signal for time series data \cite{myers2020anapt}. We explain in detail the process of filtration of time series and extraction of topological features in the next section.   

\subsection{Feature extraction using 0-D Sublevel Set Persistence}
\label{sub_level}
Feature extraction plays a crucial role in machine learning by extracting the relevant features or information from the given data (in our context, it is time series data) for the machine learning models to efficiently do the classification task \cite{kuhn2019feature}. One of the immediate consequences of using feature selection is dimensionality reduction, which inherently reduces the computational complexity \cite{khalid2014survey}. In this work, we employ sublevel set, a tool from topological data analysis, as a feature extraction mechanism to obtain meaningful features from nonlinear time series.

Sublevel set persistence deals with real-valued functions defined on a manifold and tracks the topological changes of the function as we vary the filtration parameter $c$. In simple terms, this tracks how connected components appear (birth) and merge (death) as we sweep (filter) the function or time series from the bottom to top (along the vertical direction). Let us define a function $f: \chi \to \mathbb{R}$ where $\chi$ is a topological space. For a given parameter $x \in \chi$, the sublevel set of $f$ \cite{edelsbrunner2022computational} is defined as 
\[
\chi_c = f^{-1}((-\infty, c]) = \{x \in \chi \mid f(x) \leq c \}.
\]
As $c$ increases, the sublevel sets increase to form a filtration of the topological space. The critical points of the function $f$ correspond to values of $c$ at which the topology of $\chi_c$ changes; these are called the critical events. These changes are generally characterized by the appearance of new connected components, the appearance of holes, and so on. In the case of $0-dimensional$ sublevel persistence \cite{edelsbrunner2022computational}, we deal with $1-D$ data, which is the time series data. Here, the critical points correspond to the local maxima (crests) and local minima (troughs) of the function $f$ that occur during the filtration process. We characterize a local minimum if a point has a lower value than the neighboring points, and a local maximum if a point has a greater value than the neighboring points in a time series. This filtration process allows us to sequentially track the evolution of the topological features for different height parameter values, c. 

\begin{figure*}
     \centering
     \begin{subfigure}[b]{0.50\textwidth}
         \centering
         \includegraphics[width=\textwidth]{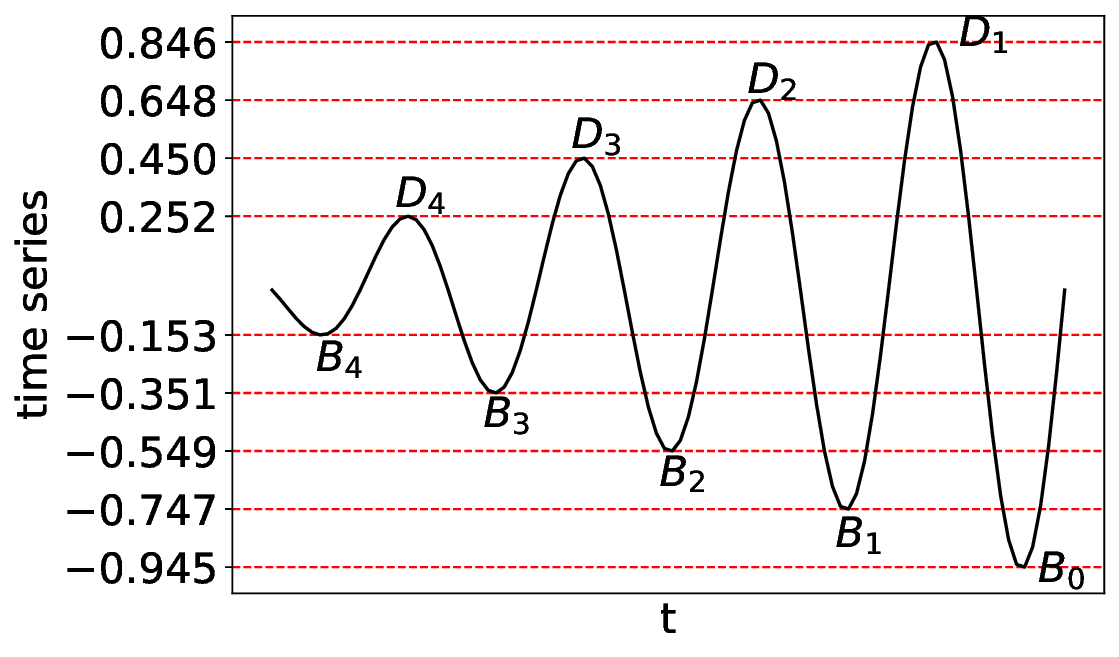}
         \caption{Filtration process of a synthetic time series data}
         \label{fig:timeseriesPD}
     \end{subfigure}
     \begin{subfigure}[b]{0.3\textwidth}
         \centering
         \includegraphics[width=\textwidth]{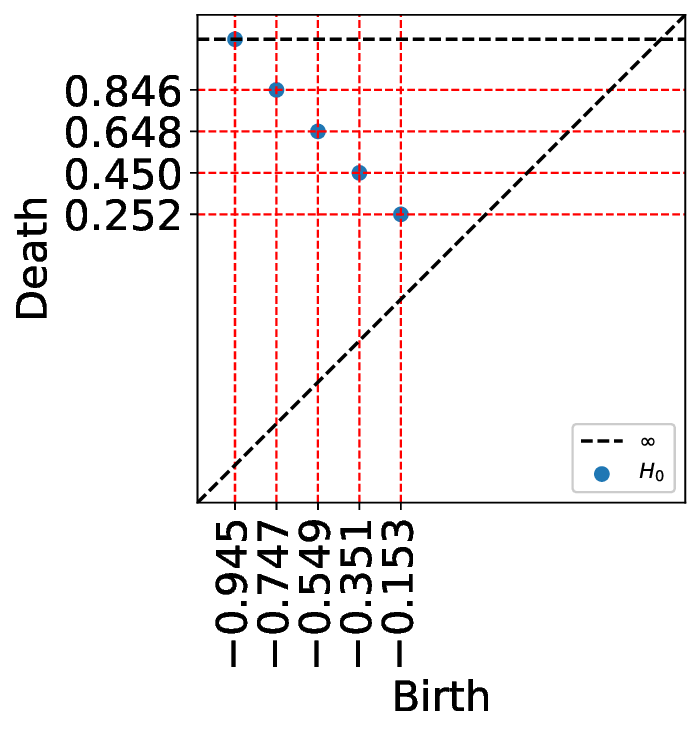}
         \caption{Sub-level set persistence diagram}
         \label{fig:demoPD}
     \end{subfigure}
     	        \caption{Figure illustrating the $0-D$ sublevel set filtration for a synthetic time series. The time series is swept along the height parameter (y-axis). The birth of a topological feature is recorded when a local minimum is encountered and its death is recorded when a corresponding local maximum is reached. The local maxima and minima are paired according to Elder's rule and represented in the persistence diagram. The time series data (figure on the left) contains five minima, denoted by $B_0, B_1, B_2, B_3, B_4$, which consequently give rise to five topological features presented in the persistence diagram (figure on the right).}
              \label{fig:timeseries_sublevel}
     \end{figure*}

We should visualize the filtration process as if the water level rises over the function along the height coordinate. Consider Fig. \ref{fig:timeseriesPD}, where we fill the function from the bottom, and water starts to fill at the first local minimum, which is denoted as $B_0$, and we increase the height c; we meet the second local minimum at $B_1$. Similarly, we mark the other minima at $B_2$, $B_3$, and $B_4$. The red dotted lines of level sets are drawn whenever a critical event occurs; specifically, when minima and maxima occur, these mark the occurrence of topological features. Now, when we sweep a bit higher, we meet the first local maximum $D_4$. Here, the disjoint valleys/pools to the left and to the right of $D_4$ merge together. Thereby, the last minimum that appeared, $B_4$ (called the birth of a topological feature) is now linked to the first maximum $D_4$ (called the death of the topological feature). Although we have five minima that were born before the occurrence of the first maximum, we associate the youngest minimum with the first maximum according to Elder's rule \cite{edelsbrunner2022computational}. Elder's Rule states that when two connected components merge (typically at a maximum), the older one persists while the younger one dies. As we go by this statement, we associate the death of the connected component that was born at $B_3$ to $D_3$, and that of $B_2$ to $D_2$, and so on. But the oldest or the first connected component that was born at $B_0$ never dies and goes to infinity, while others get paired. 

The birth and death coordinates of these topological features are presented in the form of a plot called a "Persistence Diagram" (PD) in Fig. \ref{fig:timeseriesPD} (b). This 2-dimensional plot tracks the evolution of the connected components as the height parameter \emph{c} varies. The separation of the points from the diagonal represents the persistence of these features. For example, the feature ($B_4$, $D_4$), which is ($-0.153,0.252$) in terms of c, shows the lowest persistence and is short-lived, while the feature  ($B_1$, $D_1$), which is ($-0.747,0.846$) in terms of \emph{c}, shows maximum persistence, and hence it is long-lived. 

\begin{figure*}
     \centering
     \begin{subfigure}{0.8\textwidth}
         \centering
         \includegraphics[width=\textwidth]{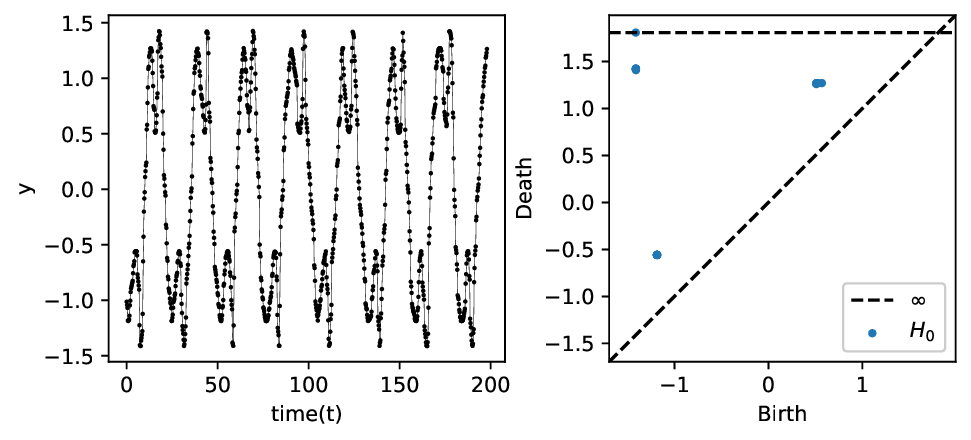}
         \caption{Persistence diagram (on the right) obtained for scanty periodic time series (on the left)}
         \label{fig:comp_35}
     \end{subfigure}
     \begin{subfigure}{0.8\textwidth}
         \centering
         \includegraphics[width=\textwidth]{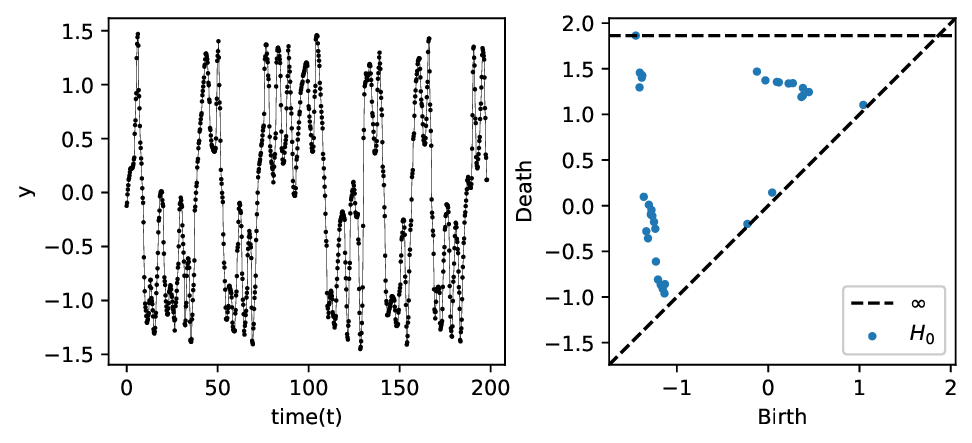}
         \caption{Persistence diagram (on the right) obtained for scanty chaotic time series (on the left)}
         \label{fig:comp_39}
     \end{subfigure}
     	        \caption{This plot is a comparison of persistence diagrams (PDs) obtained by computing sublevel set using scanty time series of $\gamma = 0.35$ (period-3 limit cycle) and $\gamma = 0.39$ (chaotic) for the Duffing system, respectively: The periodic time series data exhibits repeating, regular cycles of maxima and minima, which contribute to the overlapping, segregated topological features in PD. While the scanty chaotic time series exhibits irregular, aperiodic cycles of maxima and minima that contribute to the scattered topological features in the PD.}
              \label{fig:feature comparison}
     \end{figure*}

Now, we shall quickly demonstrate how sublevel set analysis can be applied to candidate periodic and chaotic time series data sets. In doing so, we shall use the scanty time series data sets for both of them by removing $70\%$ of the data from the time series randomly. For demonstration, we shall take a scanty periodic time series data from the Duffing system as shown in Fig. \ref{fig:feature comparison}(a). The right panel shows the persistence diagram (PD) of the periodic time series data (in the left panel) by extracting the topological features using sub-level set analysis. It is easily discernible that the few topological features in the PD are imprints of a few patterns of maxima and minima (equivalently, birth-death pairs)  that keep repeating in the time series. However, in contrast, the topological features in the chaotic regime in Fig. \ref{fig:feature comparison}(b) are seen to be scattered about. We can relate this scatter to the non-repeating maxima and minima in the chaotic time series, which helps us to demarcate it from the periodic case. 

The primary motivation for adopting a TDA-based framework lies in its robustness to missing samples and small perturbations. To understand this, let us go back to Fig. \ref{fig:feature comparison}(a), despite the removal of $70\%$ of the data samples, the important topological features remain preserved in the PD. The key takeaway is that sparsity induces only a negligible impact in the extracted topological summaries, retaining the intrinsic dynamical nature of the system. 

In summary, sub-level set analysis transforms time series data into a persistence diagram (PD) and thereby segregates the periodic and chaotic time series using their corresponding persistence diagrams. The PDs so obtained contain the distilled summaries from the time series of both periodic and chaotic cases. The PDs of the periodic cases display only a very few point clusters that are tightly knit with very little dispersion. But on the other hand, the PDs of chaotic cases are spread out into many clusters that are, in turn, dispersed. The task now is to convert these topological features into a form that is feasible to feed into the ML algorithm. To do this, we stick to a process called "vectorization".

\subsection{Vectorization of Persistent Diagrams} 
\label{vectorization}
As stated earlier, persistence diagrams represent topological features as data points in persistence diagrams (PDs). These are nothing but the list of (birth, death) coordinate pairs of the topological features. Consequently, PDs extracted from different time series contain varying numbers of topological features or feature points, even when the time series considered are of equal length and sampled at the same frequency. This variability primarily arises not because of the sampling parameters; instead, it is determined by qualitative dynamics -- by this, we mean the number of minima and maxima that the time series signal exhibits.

To put things into perspective, the bifurcation ranges considered in this study encompass both periodic and chaotic regimes. Even within the periodic dynamics, we have time series signals varying limit cycles (hallmark of bifurcation). For instance, in the case of the Duffing system, the system transitions through multiple periodic states, such as period-3, period-4, and period-5 limit cycles in the periodic regime (from $\gamma=0.35$ to $\gamma=0.38$) and then eventually transitions into the chaotic regime. All these dynamical states (including the chaotic signals) exhibit a different number of extrema, which in turn produces varying numbers or counts of topological features. Consequently, this is reflected in the number of topological feature points that occur in the PD. The immediate consequence is that the raw topological features cannot be fed into the machine learning (ML) classifiers since most of the ML models require the inputs to be represented as fixed-length feature vectors. 

To counter this issue, \emph{featurization} or \emph{vectorization} of persistence diagrams becomes a necessary preprocessing step. Vectorization transforms the variable-length topological feature points contained in the diagrams into a structured numerical representation, known as feature vectors, which meets the requirements of machine learning algorithms. This process involves computing descriptive statistics or structured transforms that capture the topological information encoded in the PDs. Consequently, in the literature \cite{ali2023survey,khasawneh2018chatter}, Carlsson coordinates \cite{adcock2013ring} have been used to featurize or vectorize the topological points in the PDs, and hence we stick to the same protocol in our work. In addition to these four coordinates, we propose another coordinate that computes the variance of the pairwise Euclidean distances between points in the PD. Hence, for a list of topological feature points presented in a PD, $\mathcal{D} = \{(x_1, y_1), (x_2, y_2), (x_3, y_3) \dots, (x_n, y_n)\}$, the five input feature vectors are defined as follows:
\[
f_1 (\mathcal{D}) = \sum_i x_i (y_i - x_i)
\]

\[
f_2(\mathcal{D}) = \sum_i (y_{\text{max}} - y_i)(y_i - x_i)
\]
\[
f_3(\mathcal{D}) = \sum_i x_i^2 (y_i - x_i)^4
\]
\[
f_4(\mathcal{D}) = \sum_i (y_{\text{max}} - y_i)^2 (y_i - x_i)^4
\]
\[
f_5(\mathcal{D}) = \mathrm{Var}\left( \left\| (b_i, d_i) - (b_j, d_j) \right\| \right)
\]

Where $x_i$ and $y_i$ are the birth and death of the topological feature points, respectively and $y_{max}$ is the maximum death value. Each input feature has a purpose and we state them below:

\begin{enumerate}
    \item $f_1 (\mathcal{D})$ is the birth-weighted lifetime (of a topological feature). While the lifetime $(y_i - x_i)$ carries the information of how long each feature persists, the birth weight incentivizes the topological features that appear early in the filtration.
    
    \item $f_2(\mathcal{D})$ measures the difference between the maximum death value and the death time of the feature, multiplied by its lifetime. The term $(y_{\text{max}} - y_i)$ highlights the topological features that die early during the filtration.    

    \item $f_3(\mathcal{D})$ represents the higher-order extension of $f_1 (\mathcal{D})$. This feature amplifies the contribution of long-lived or dominant topological features while reducing the influence of the short-lived or noisy topological features. 
    
    \item $f_4(\mathcal{D})$ also represents the higher-order extension of $f_2(\mathcal{D})$. The emphasis or amplification is given to the lifetime or persistence of the features through higher-order terms. The consequence is to enhance the contribution of long-lived or persisting features that preserve the global topology of the time series. 

    \item $f_5(\mathcal{D})$ represents the variance of pairwise Euclidean distances between the topological features. This is a measure of dispersion of the topological features and quantifies the distributional structure. In addition to the four Carlsson coordinates ($f_1-f_4$), we introduced this descriptor to demarcate the topological features that cluster and scatter. 
\end{enumerate}

Collectively, these features all provide a compact statistical representation and description of topological features. As highlighted earlier (in Fig. \ref{fig:feature comparison}), these are designed to effectively discriminate between the tightly clustered features arising from the periodic dynamics and the scattered features arising from the chaotic dynamics. We shall now discuss how the ML classifiers can be used for this purpose in the next section.

The sublevel set visualizations shown in the paper, and the computations were performed using the RIPSER library from scikit-tda \cite{scikittda2019} in Python. The computations were run on a computer with an Intel(R) Core(TM) i5-8300H processor with a CPU speed of 2.30 GHz using 8.00 GB of RAM running on Windows 11.

\begin{figure}
         \centering
         \includegraphics[width=0.48\textwidth]{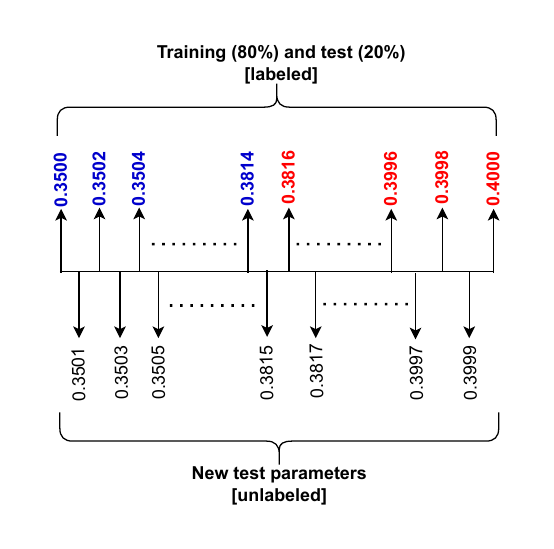}
         \caption{Graphical representation illustrating the selection process of parameters for the training set and test set (parameters at the top), which are labeled, along with the \emph{new test parameters} (parameters at the bottom), which are unlabeled, from the entire pool of available Duffing system bifurcation parameters ($\gamma$). In the figure, blue-colored parameters denote that they are from the periodic regime, while red-colored parameters represent those associated with the chaotic regime. As shown, the sets of parameters used for the training set and the test set are non-overlapping with those used for \emph{new test parameters}, thereby giving a fair assessment of the model's generalizability.}
         \label{fig:param_selection}
\end{figure}

\section{Machine Learning Classifier}
\label{sec:ML}
Machine learning (ML) classifiers have become potential tools across various domains, including the analysis of fMRI data \cite{pereira2009machine}, malware detection in mobile devices \cite{narudin2016evaluation}, seizure detection from EEG signals \cite{siddiqui2020review}, and diabetes prediction \cite{hasan2020diabetes}. These ML classifiers are useful, particularly when the target variable is categorical, enabling classification or detection based on the input features. In our study, we focus on classifying time series as either periodic or chaotic, leading to a binary classification problem. Broadly, ML approaches can be classified into supervised learning and unsupervised learning. Supervised learning relies on labeled input data to learn the patterns and then categorizes the unseen data. On the other hand, unsupervised learning does not rely on labeled input data; instead, it autonomously identifies inherent patterns and commonalities within the data and groups them accordingly. In this work, we focus on supervised learning models, as labeled data are available or can reasonably be obtained.

\subsection{Classifier Models}
\label{model}
As mentioned in Sec. \ref{sub_level}, the topological feature pairs from the scanty time series are extracted by computing sublevel set homology and vectorized using the five input feature vectors. We do this exercise for a range of bifurcation parameter values, encompassing both periodic and chaotic regimes. We label the input features manually: those obtained from periodic time series are labeled as 0, and those obtained from chaotic time series as 1. These labeled features serve as input data for the supervised ML classifiers. 

We illustrate this process using the scanty time series data obtained from the Duffing system. To mimic the real-world scenarios, where one often has to deal with data that is scanty or laden with missing elements (data points), we intentionally simulate a regularly sampled time series data and randomly remove data points from it. It is important to note that throughout this study, the scanty or sparse time series data were generated using a \emph{uniform random deletion} process. By this, we mean that the data points were removed randomly across the time series without imposing any structural or consecutive missing patterns. For the Duffing system, we simulate a $198s$ long time series data and consider it as the unperturbed signal with 3000 data points. To make it scanty, we randomly remove $70\%$ of the data, which amounts to 2100 data points, leaving a sparse time series consisting of 900 data points. We then repeat the same process for bifurcation parameters ranging from $\gamma=0.3500$ to $\gamma=0.4000$, in steps of 0.0002. We compute sublevel set homology for each time series to extract its topological features and vectorize it. Based on known bifurcation behavior, the input features extracted from time series up to parameter value $\gamma=0.3814$ are considered periodic and labeled or tagged as `0', while those from $\gamma=0.3816$ to $\gamma=0.4000$ are considered chaotic and labeled as `1'. These labeled and vectorized features are compiled into a single master dataset to train and validate the supervised machine learning classifiers. 

Generally, in ML algorithms, the labeled or annotated master dataset is shuffled and split into three sets, viz., training, validation, and test sets. The training set is where the model learns the underlying pattern and computes a threshold to classify the unseen data into the respective labels or classes (in our study, this classification task involves distinguishing the periodic and chaotic regimes). The validation and test sets are reserved for model tuning and final performance assessment, respectively \cite{burkov2019hundred}. While there is no fixed proportion to split the data, the rule of thumb is to dedicate a larger portion to the training set in order to maximize the learning. The common proportion that is used to split the dataset is the following: $70\%$ for
the training set, $15\%$ for the validation set, $15\%$ for the test set \cite{burkov2019hundred}. However, in situations where the available annotated or labeled data is limited, the validation set can be drawn from the training data using cross-validation techniques \cite{burkov2019hundred}, as discussed in the next section. This approach avoids the need for a separate validation subset. But the role of the validation set is critical, as it helps the model choose the optimal set of hyperparameters \footnote{In ML, hyperparameters are parameters that can be configured for the learning process in order to extract the full potential of a model in classifying particular data under study}. While the role of the test set is to evaluate the model's performance when it is presented with hitherto-unseen data. The results of this exercise are measured by some standard metrics that ensure that the model is not overfit and can reliably perform in real-world applications.

Now, let us discuss in detail the models employed in the work. We choose three widely used supervised learning binary classifier models: Logistic Regression (LR), k-Nearest Neighbors (kNN), and Support Vector Machines (SVM). 

\begin{enumerate}
    \item \textbf{LR classifier:} LR classifiers are generally preferred when we deal with a binary classification problem. It relates the independent variables, or input features, $x_1, x_2, \ldots, x_n$ to the target variable (y) using an `S' shaped curve called a Sigmoid or Logistic function \cite{zhou2021machine}. The sigmoid function maps any real-valued input to a value between 0 and 1. We define the sigmoid function as $\sigma(y) = \frac{1}{1 + e^{-y}}$ where $y = b + w_1 x_1 + w_2 x_2 + w_3 x_3 \ldots + w_n x_n$. Here, $w_1,w_2,w_3,\ldots,w_n$ are the weights associated with the respective input features, and $b$ is the bias term. The outcome of the sigmoid function represents the predicted probability of an input belonging to a particular binary class, either 0 or 1. Generally, a threshold value of $0.5$ determines the binary class or category it falls under. If the outcome value is below that threshold value, it will result in class 0, and if the outcome value is above, it will result in class 1. 

    \item \textbf{SVM classifier:} It is a powerful tool capable of performing linear and non-linear classifications. This algorithm \cite{burkov2019hundred} aims to find an optimal hyperplane in $N$-dimensional space that separates the datasets into their respective classes; in our case, it is periodic and chaotic. The optimal hyperplane is the one that maximizes the margin, which is the distance between the closest points of either class. These closest points are called support vectors, and they play a major role in determining the position and orientation of the hyperplane. The number of input features determines the dimension of the hyperplane; the general notion is that for $N$ input features, the hyperplane exists in $N-1$ dimensions. However, in places where the data is not separable in its original space, practitioners can opt to change the kernel function, which is a hyperparameter for this model. The kernel function maps or transforms input data to a higher-dimensional space, such that it is feasible for an optimal hyperplane separation \cite{zhou2021machine}. Commonly used kernel functions include the Radial Basis Function (RBF) or Gaussian kernel, linear kernel, and polynomial kernel. 

    \item \textbf{kNN classifier:} It is one of the simplest and most versatile algorithms. The widespread use of this algorithm is attributed to its non-parametric approach, by which we mean that it does not learn any fixed parameters (like weights and bias) during training. The algorithm \cite{burkov2019hundred} identifies the `k' nearest neighbor points (which is a hyperparameter) for a data point based on the chosen distance metric. kNN is often termed a "lazy learner" since, unlike other ML algorithms, it explicitly does not build a model for the training set. Instead, it relies on storing the feature vectors with their respective labels and performs classification for a new data point or test set points only when they are encountered. During the testing phase, when a new unlabeled data point is presented, it finds the `k' neighboring points of the training data in terms of distance. The class label is assigned by taking the majority of the labels of the `k' neighboring points. 
\end{enumerate}
 
We used the above-mentioned three classifier models and compared their performances in order to identify the most efficient model for a given system, to help the practitioner choose the better model. We ensure an identical data set comprising the test set ($20\%$ of data) is used during the evaluation process on all three models so that they are evaluated on the same grounds, and the metrics derived from them can be compared under identical conditions. This is done by retaining the same random seed during the data splitting step. The split proportion was kept random (not stratified), as the class distribution in the labeled dataset is largely balanced. The specific evaluation metrics used for performance comparison are discussed in the next section. 

The ML computations in this paper were carried out using the scikit-learn \cite{sklearn_api} in Python. The calculations are performed with the same system configurations specified in the previous section.

\begin{figure*}
         \centering
         \includegraphics[width=\textwidth]{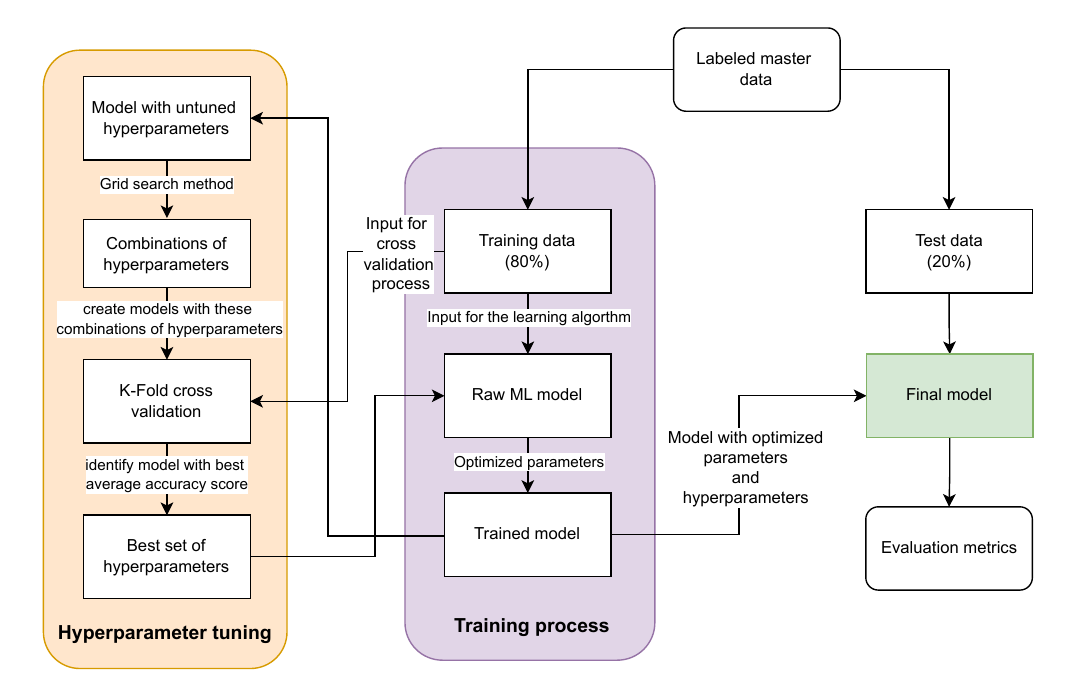}
         \caption{Machine learning workflow.}
         \label{fig:ML_flowchart}
\end{figure*}

\subsection{Evaluation Metrics and Hyperparameter Tuning}
\label{metrics}
Evaluation metrics are required to evaluate and grade the performance of ML models before they are released for deployment. These metrics are computed on an independent \emph{test set}. The evaluation metrics compare and quantify the agreement between predicted and actual labels \cite{marsland2011machine}. In this study, performance is assessed using four important key metrics: accuracy, precision, recall, and F1 score, which are derived from the confusion matrix \cite{raschka2019python}.

Let us quickly outline how evaluation metrics are calculated. The summary of correctly and wrongly classified outcomes of the ML model run on the test dataset is presented in the form of a matrix, called a confusion matrix. It is a $K \times K$ matrix \cite{raschka2019python}, where $K$ is the number of classes or target labels, which in our case is $K=2$ since it is a binary classification problem, which leads to four elements in the matrix. The four elements are true negatives (TN) where the model's prediction is 0 and the actual label is 0; false positives (FP) where the model's prediction is 1 and the actual label is 0; false negatives (FN) where the model's prediction is 0 and the actual label is 1; and true positives (TP) where the model's prediction is 1 and the actual label is 1. We shall use the four most commonly used evaluation metrics \cite{marsland2011machine} in machine learning here and provide their definitions  below: 

\begin{enumerate}
    \item \textbf{Accuracy:} It is defined as the ratio of the number of correct predictions to the total number of input samples. This gives us an idea about the model's performance in predicting the correct labels. It is defined as
    \[
     Accuracy = \frac{TP + TN}{TP + FN + TN + FP}.
    \]
    \item \textbf{Precision:} Precision measures how many positive predictions that were made by the model are truly positive or correct. Precision is significant when false positives play a crucial role, and is given by,
    \[
     Precision = \frac{TP}{TP + FP}.
    \]
\item \textbf{Recall:}
    Recall is more of a sensitivity measure, which determines how many correct positive observations were made from all the actual positive observations. Recall is significant when false positives play a crucial role. It is defined as:
       \[
     Recall = \frac{TP}{TP + FN}.
    \]
\item \textbf{F1 Score:}
    F1 score is the harmonic mean of \emph{precision} and \emph{recall}, which provides the right weightage between both, and is defined as:
    \[
     F1 Score = \frac{2*precison*recall}{precision +recall}.
    \]
\end{enumerate}

Evaluation metrics can help us diagnose and resolve issues like overfitting and underfitting. Furthermore, comparing the computed values of the evaluation metrics of different models will give us a fair idea about the suitability and effectiveness of the models under different conditions. This helps the practitioner pick the appropriate model for specific tasks. In addition, it is imperative to note that the model performance is influenced not only by the learned parameters (during the training process) but also by a set of manually specified \emph{hyperparameters} \cite{bischl2023hyperparameter}. Consequently, it is essential to find the appropriate hyperparameters for achieving the optimal classification accuracy. In our work, we employ the grid search method \cite{marsland2011machine}, which is the method of exhaustion, where we first create all possible sets of combinations of the hyperparameters using the discrete values for each of the hyperparameters.

To avoid overfitting and ensure unbiased model selection, we employ $K$-fold cross-validation on the training data during hyperparameter tuning. This technique allows one to have just two major splits of the master dataset: the training set and the test set. In our work, we choose $80\%$ for training and $20\%$ for testing. The training set is further split into $K$-folds, where $K$ is a finite number. Here, ($K-1$) folds are used for training, and the remaining $1$-fold is used for validation, where its accuracy is determined. This process is repeated $K$ times, and the average validation accuracy is determined. The combination of hyperparameters that produces the highest average accuracy score is identified as the optimal hyperparameter configuration \cite{burkov2019hundred}. Once the optimal hyperparameters are chosen, the final performance is then evaluated on the independent (reserved) test set, which remained untouched throughout the hyperparameter tuning process. The philosophy of this approach is that it ensures hyperparameter optimization, model training, and model performance evaluation remain separated. 

The overall workflow is depicted in the flowchart in Fig. \ref{fig:ML_flowchart}, and a summary of the procedure is provided as follows:

\begin{itemize}
    \item We split the master labeled dataset into two sets: one with $80\%$ for the training set and the remaining $20\%$ for the test set to evaluate the performance of the ML model.
    \item We need to tune the hyperparameters to extract the best performance of the model, and hence we use the grid search method to create a set of combinations of the hyperparameters. 
    \item We need a validation set to evaluate the performance (or the accuracy score) of the model for the generated set of combinations of hyperparameters, for which we use K-fold cross-validation -- A technique that generates validation sets from the training set (Note: $20\%$ test set is still left unperturbed) and computes the accuracy score for the validation sets.
    \item The combination of hyperparameters that produces the best accuracy score is chosen as the model's optimal hyperparameters.
    \item In the final stage, the evaluation metrics are obtained by assessing the ML model's performance on the test set, where the ML model's hyperparameters were tuned to their optimal values. 
\end{itemize}

However, we do not stop with the test set data. We further put our trained models (LR, SVM, kNN) under test with new bifurcation parameters that were not exposed to the model during the training and test set, as illustrated in Fig. \ref{fig:param_selection} for two major reasons. First, to check whether the models can perform a real-world distinction between periodic and chaotic data. Second, to benchmark the performance of the classifiers against well-established traditional indicators, such as the maximum Lyapunov exponent (MLE), in order to validate their consistency and robustness. We term these parameters as \emph{new test parameters}. For the \emph{new test parameters} set (we use the same Duffing system for explanation purposes), we construct an independent set by generating scanty time series for a non-overlapping range of parameters, $\gamma=0.3501$ to $\gamma=0.3999$ in steps of 0.0002. This process is illustrated in Fig. \ref{fig:param_selection}, where we show how the labeled data are split (we highlight the parameters that exhibit periodic behavior in blue and the ones that exhibit chaotic behavior in red) into the training and the test set (look at the range of parameters that are placed at the top) while the non-overlapping parameters that are unlabeled are taken as \emph{new test parameters} (look at the range of parameters that are placed at the bottom). 

This approach ensures that it sweeps similar dynamics, but the parameters are entirely distinct from the training, validation, and test sets used. We then extract the topological features by computing sublevel set homology and allowing the ML classifiers to classify them. We deliberately leave the \emph{new test parameters} unlabeled to perform a blind comparison against traditional dynamical systems techniques, particularly the maximum Lyapunov exponent methods. This allows us to showcase how well our trained ML classifiers perform on unseen data (by classifying the \emph{new test parameters} tag 0 if they are periodic and 1 if they are chaotic).

 Broadly speaking, one would notice that these fully trained, validated, and tested classifiers (LR, SVM, kNN) typically function like binary score quantifiers for the \emph{new test parameters}, assigning discrete labels of 0 or 1 (with no intermediate values -- no grey area), which also serves as a strength for our approach. This makes us associate our quantifiers with the $0-1$ chaos test \cite{gottwald2004new}, where their methodology classifies the time series into 0 for periodic and 1 for chaotic. We benchmark our ML-classified binary scores against the maximal Lyapunov exponent to ascertain the predictions made by the classifiers.

\section{Results}
\label{sec:Results}
We choose three well-known systems in nonlinear dynamics for our study, where we use the ML models to detect the state change when the system parameter is varied.  The three systems are the 2-dimensional Duffing oscillator, the 3-dimensional systems, the Lorenz attractor, and the Rossler attractor. Along with that, we also apply the method to real-world ECG data. These systems were chosen to demonstrate the versatility of our algorithm across 2-dimensional and 3-dimensional dynamical nonlinear systems. Further, we also apply our method to real-world ECG data to classify healthy and unhealthy cases, and thereby validate our claim.

\subsection{Duffing system}
\label{duff_results_section}
The Duffing oscillator exemplifies a two-dimensional, non-autonomous, and non-linear damped-driven system. It demonstrates a period-doubling cascade and chaotic behavior due to the presence of non-linear terms. The equation governing its motion is given by:
$$
\ddot{y}+\delta \dot{y}+\alpha y(t)+\beta y^{3}(t)=\gamma \cos (\omega t)
$$
where, $\gamma$ represent the driving force amplitude and $\delta$ represents the damping coefficient. The bifurcation route to chaos occurs by changing the bifurcation parameter, gamma $(\gamma)$, and keeping other parameters constant $\beta=1$, $\alpha=-1$, $\delta=0.3$, $\omega=1.2$, $y(0)=0$, $\dot{y}(0)=0$. 

\begin{figure}
     \centering
     \begin{subfigure}[b]{0.45\textwidth}
         \centering
         \includegraphics[width=\textwidth]{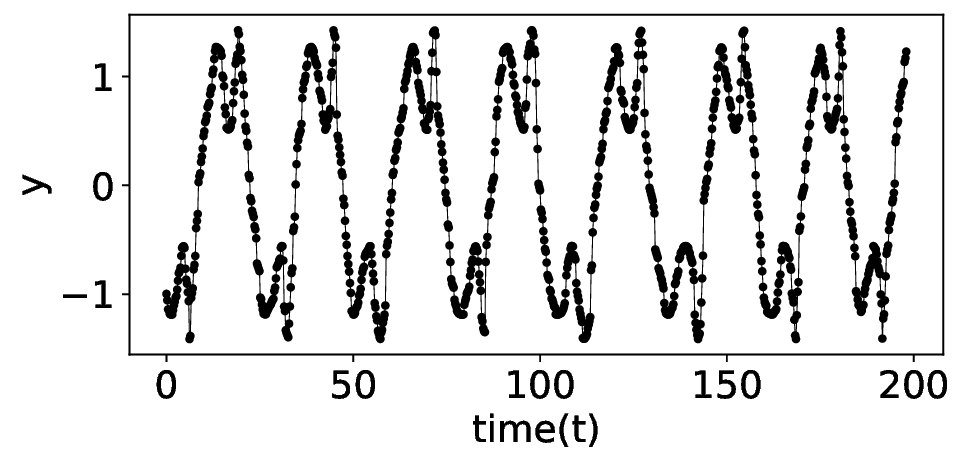}
         \caption{}
         \label{fig:duff_35}
     \end{subfigure}
	 \begin{subfigure}[b]{0.45\textwidth}
         \centering
         \includegraphics[width=\textwidth]{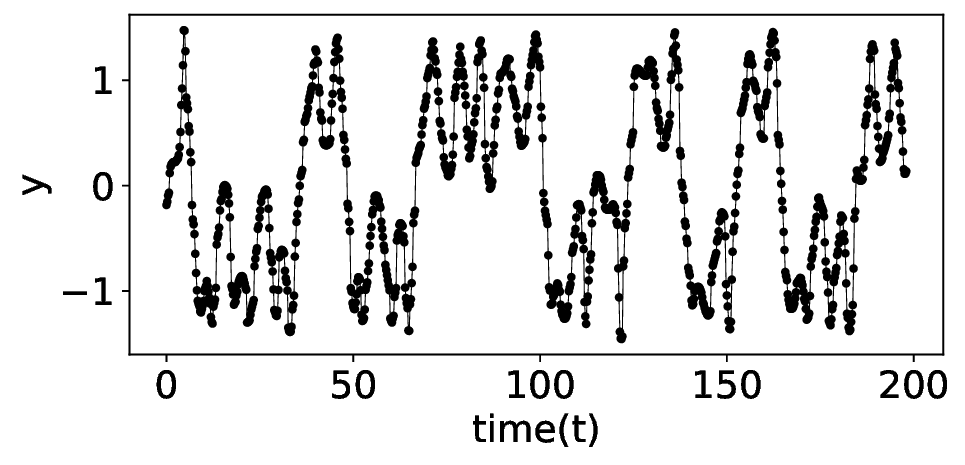}
         \caption{}
         \label{fig:duff_39}
     \end{subfigure}
     	        \caption{Scanty time series data of the Duffing system exhibiting (a) periodic behavior at  $\gamma=0.35$ and (b) chaotic behavior at $\gamma=0.39$.}
        		\label{fig:duff_bifurcation}
     \end{figure}

The system is evolved over a time interval of $t=0$ to $t=210\pi$ seconds, which is approximately $660$ s, with a step size of $0.066$ s. This corresponds to a sampling frequency of $15.2$ Hz. We extract the last $198$s  window consisting of 3000 data points as the representative signal or time series data. As stated earlier, to mimic the real-world scenarios wherein the time series often contain missing elements, we randomly remove $70\%$ of the data  (which is equivalent to 2100 data points), leaving a scanty time series containing 900 data points. We do this exercise keeping two things in mind. i) Data getting compressed/reduced during processes; ii) The data acquired itself was limited. That is, by omitting random data from the complete dataset, we aim to mimic processes that mimic how partial and scanty data are generated/recorded in experiments and real-world data. In Fig. \ref{fig:duff_bifurcation}, we illustrate examples of such scanty time series for the Duffing system at two different values of the bifurcation parameter ($\gamma$): One at $\gamma=0.35$ for demonstrating periodic behavior and another one at $\gamma=0.39$ for the chaotic behavior. 

Now we shall briefly explain how the master dataset is prepared for the ML analysis using the three ML models. 
We compute the sublevel set homology for each scanty time series extracted over a range of bifurcation parameters from $\gamma=0.35$ to $\gamma=0.40$ in steps of 0.0002 (which accounts for $251$ parameter instances in total). That is, there are $251$ scanty time series data in total that got extracted at $251$ parameter instances, which include both periodic and chaotic time series.  

Next, we extract the topological features of each of these scanty time series data and follow the same procedure mentioned in Sec. \ref{model}. The extracted topological features are vectorized into 5 input features ($f_1,f_2,f_3,f_4,f_5$) following the procedure mentioned in Sec. \ref{vectorization} for all the $251$ time series and then compiled into a single master dataset.
The target class for the input features extracted for each scanty time series data (at a bifurcation parameter) is manually labeled as category '0' if they are \emph{periodic} and classified as category '1' if they are \emph{chaotic}, based on the known dynamical behavior. Now, this master data set is ready for use. As we discussed, we use $80\%$ of the master dataset as the training set (200 samples), and $20\%$ of the master dataset as the test set (51 samples) to evaluate the performance of the three classifier models.

As mentioned earlier, we use three classifiers: LR, SVM, and kNN to find the model that is best suited for the task. We use the same $80\%-20\%$ training-test split ratio for all the classifiers and keep a constant random seed for the test set (a parameter that controls the randomization). This ensures that the test set data used for the three classifiers is the same, and their performance metrics can be judged on the same grounds. The three models learn the patterns from the $80\%$ training set and compute a threshold that classifies unseen data into the respective classes. Subsequently, these models are evaluated using $20\%$ test set data to assess their learning capabilities, and the results are presented using the performance metrics mentioned above in Sec. \ref{metrics}.

\begin{figure}
\centering
\begin{minipage}{.5\linewidth}
\centering
\subfloat[Confusion matrix for LR]{\label{fig:cnf_LR_duff}\includegraphics[scale=.22]{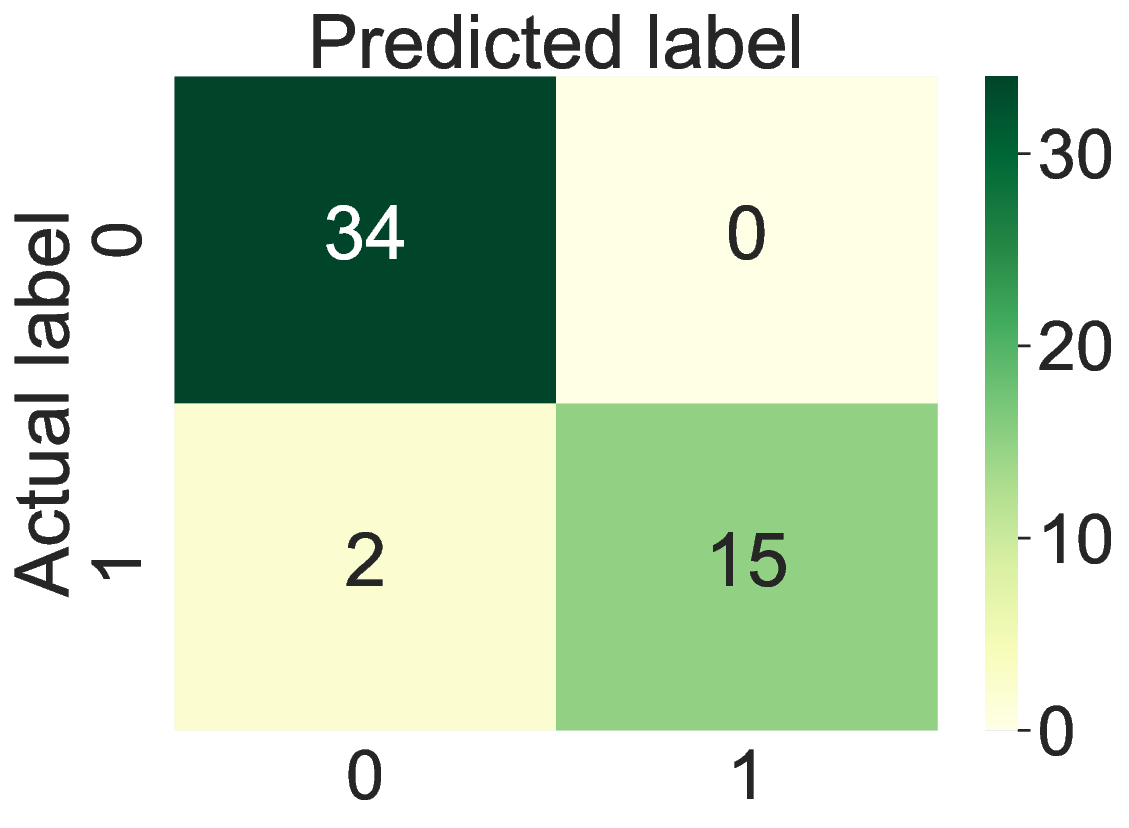}}
\end{minipage}%
\begin{minipage}{.5\linewidth}
\centering
\subfloat[Confusion matrix for SVM]{\label{fig:cnf_SVM_duff}\includegraphics[scale=.22]{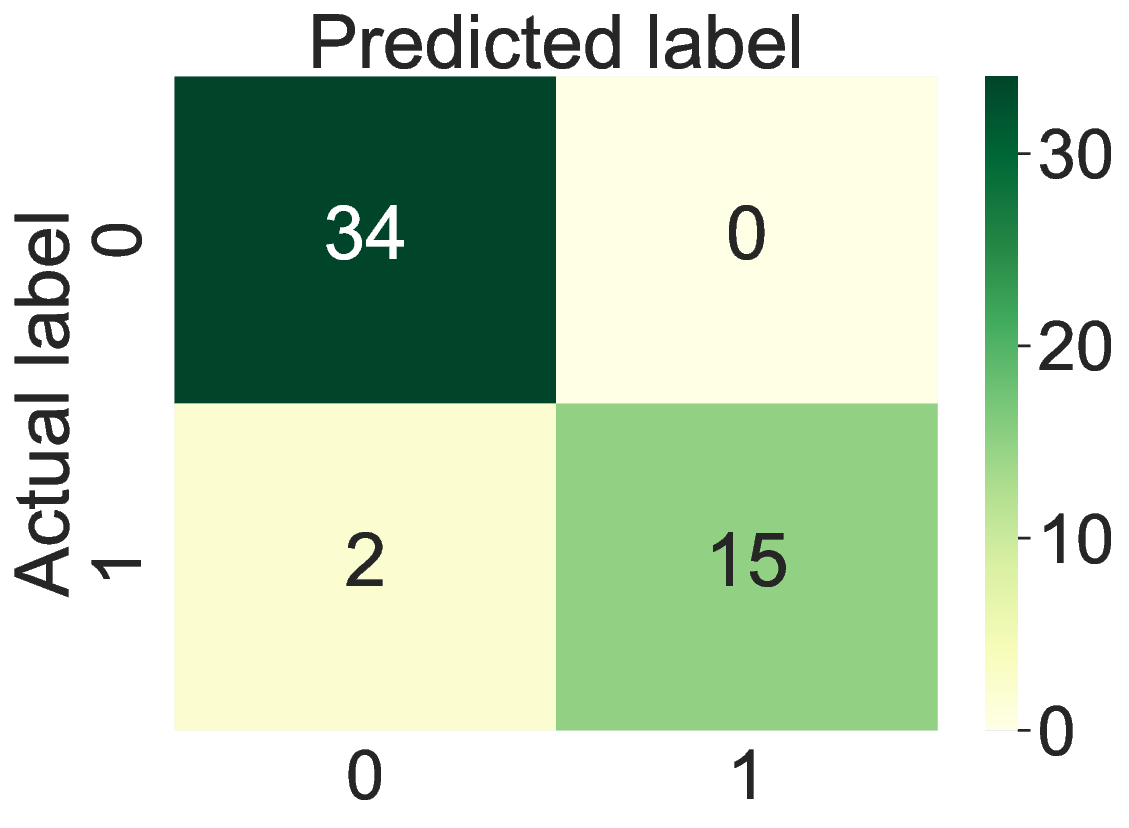}}
\end{minipage}\par\medskip
\centering
\subfloat[Confusion matrix for kNN]{\label{fig:cnf_kNN_duff}\includegraphics[scale=.22]{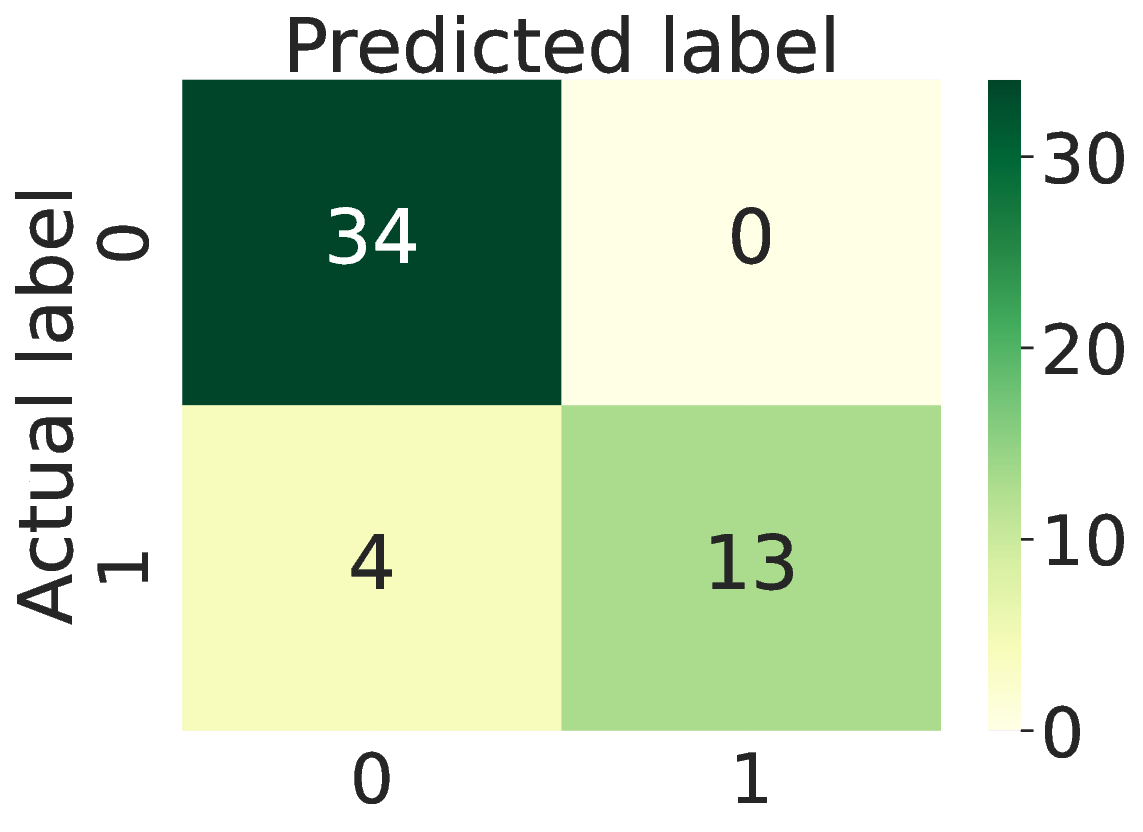}}
\caption{This figure compares the confusion matrix of LR, SVM, and kNN for the Duffing system obtained by evaluating on the same set of test data samples. The test set size accounts for 51 data samples; both SVM and LR correctly predicted 49 samples, while kNN correctly predicted 47 samples. We have a total of 4 misclassifications from kNN and 2 from both SVM and LR.}
\label{fig:cnf_mat_duff}
\end{figure}

\begin{figure}
\centering
\includegraphics[width=0.5\textwidth]{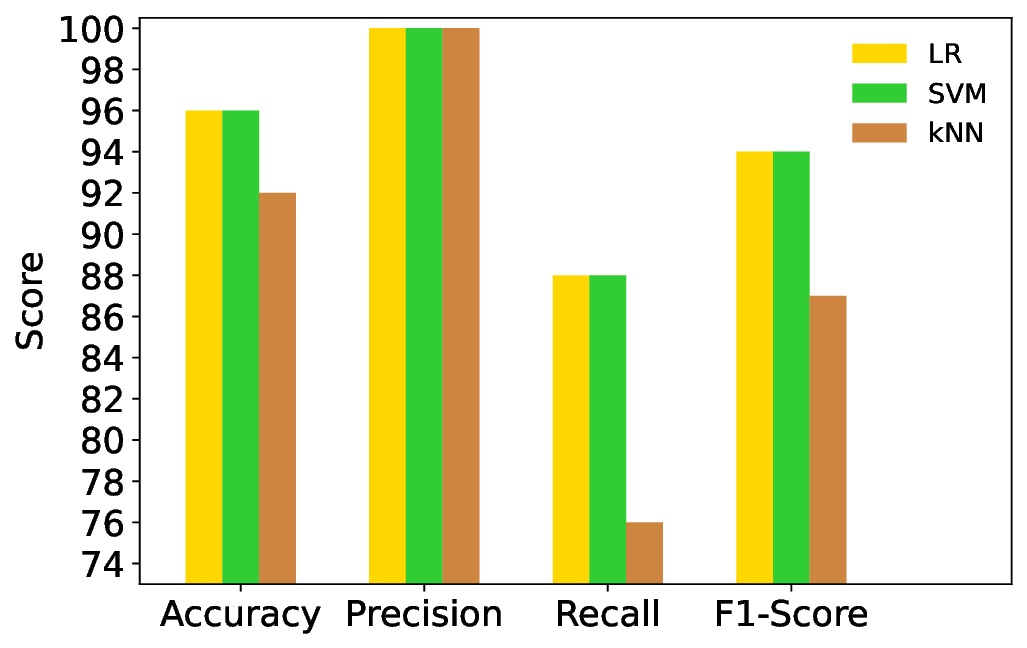}
    \caption{We present the evaluation metric scores for LR, SVM, and kNN on the Duffing system. They all obtained excellent scores; however, SVM and LR seem to be the best classifiers, while kNN seems to be on the lower side compared to their rivals.}
    \label{fig:metrics_duff}
\end{figure}

The confusion matrix for the three classifiers is shown in Fig. \ref{fig:cnf_mat_duff} to showcase the correctness in the prediction of categories. Here, all three classifiers: LR, SVM, and kNN, have correctly predicted 34 samples as periodic, while kNN correctly predicted 13 samples, and both SVM and LR have correctly predicted 15 samples each as chaotic. Furthermore, both LR and SVM have wrongly predicted 2 samples, and kNN has wrongly predicted 4 samples each as periodic. While all three: LR, SVM, and kNN have not wrongly predicted the periodic samples. This shows that LR and SVM have the least number of wrongly predicted samples, which is 2; however, let us look at the metric scores to decide the best-performing classifier for this system. We plot the evaluation metric scores in Fig. \ref{fig:metrics_duff}. As the classification results from the confusion matrix suggested, we see that both LR and SVM showed excellent scores in all four departments, while kNN seems to be underperforming compared to the other two. Additionally, the performance of the classifiers was evaluated over 30 independent random seeds, and the average accuracy obtained for LR is $97.84\% \pm 1.92\%$, SVM is $97.52\% \pm 1.75\%$, and for kNN is $90.92\% \pm 3.60\%$.

The hyperparameters of the classifiers were tuned to achieve the metric scores. The tuned hyperparameter details for all three classifiers are listed in Table \ref{tab:hyp_para_duff} to ensure reproducibility. We present the K-fold cross-validation scores to showcase that the hyperparameters chosen produce excellent and consistent accuracy scores for different validation sets (in each trial). We chose $K = 5$, where the process is repeated for 5 trials (with each trial having a different fold as the validation set), and we achieved an average accuracy score of $99\%$ for LR, $98.5\%$ for SVM, and $90.5\%$ for kNN. We tabulate the accuracy scores for each trial in Table. \ref{tab:table_duff_acc} along with its average accuracy score. The obtained high average accuracy scores certify that the hyperparameter choices are optimal and the models generalize well for the unseen data.

\begin{table*}
\centering
\begin{tabular}{|ll|ll|ll|}
\hline
\multicolumn{2}{|l|}{\textbf{LR}}                                                 & \multicolumn{2}{l|}{\textbf{SVM}}                                       & \multicolumn{2}{l|}{\textbf{kNN}}                                    \\ \hline
\multicolumn{1}{|l|}{\textbf{Hyper-parameter(s)}}          & \textbf{Value(s)}    & \multicolumn{1}{l|}{\textbf{Hyper-parameter(s)}} & \textbf{Value(s)}    & \multicolumn{1}{l|}{\textbf{Hyper-parameter(s)}} & \textbf{Value(s)} \\ \hline
\multicolumn{1}{|l|}{\multirow{3}{*}{Maximum   iteration}} & \multirow{3}{*}{200} & \multicolumn{1}{l|}{Kernel   function}           & linear               & \multicolumn{1}{l|}{No. of   neighbors (k)}      & 3                 \\ \cline{3-6} 
\multicolumn{1}{|l|}{}                                     &                      & \multicolumn{1}{l|}{\multirow{2}{*}{C}}          & \multirow{2}{*}{100} & \multicolumn{1}{l|}{Weights}                     & distance          \\ \cline{5-6} 
\multicolumn{1}{|l|}{}                                     &                      & \multicolumn{1}{l|}{}                            &                      & \multicolumn{1}{l|}{Distance   metric}           & Manhattan         \\ \hline
\end{tabular}
\caption{Table representing hyperparameter details for LR, SVM, and kNN for the Duffing system.}
\label{tab:hyp_para_duff}
\end{table*}

Using the trained models of LR, SVM, and kNN with their optimal hyperparameters, we extend the periodic and chaotic prediction of different dynamical states for a non-overlapping range of bifurcation parameters (as shown in Fig. \ref{fig:param_selection}), and term it as \emph{new test parameters}. We would like to reiterate that we intentionally use a different set of bifurcation parameters (by considering the intermediate parameters) that were not used in the training and test sets to showcase the predictability and versatility of these models. This task is done to check whether these models can distinguish the chaotic phase of the system from all the new bifurcation states if they are trained with different samples. Now, we consider scanty time series of bifurcation parameters from $\gamma=0.3501$ to $\gamma=0.3999$ in steps of 0.0002 as \emph{new test parameters}, and then we compute the sublevel set homology for these time series, and then vectorize them. Now, these vectorized input features corresponding to each bifurcation parameter value are sent into our models for classification. Now, our trained models will act as binary quantifiers that distinguish periodic and chaotic time based on the features extracted. 
 
\begin{figure}
\centering
\includegraphics[width=0.5\textwidth]{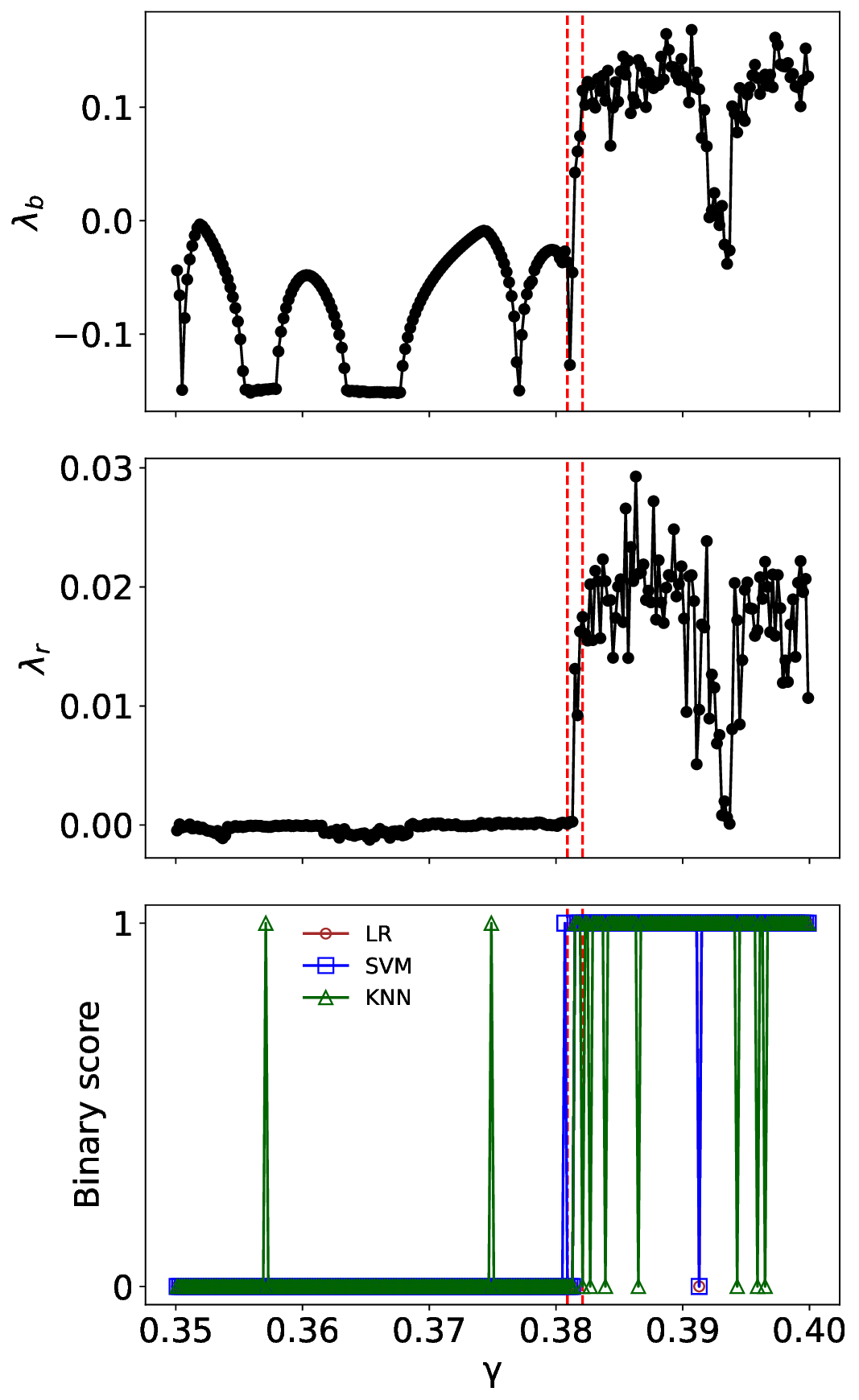}
    \caption{We showcase the classification made by the trained classifiers of LR, SVM, and kNN using the binary scores for \emph{new test parameters} of the Duffing system, where 0 connotes a periodic state while 1 connotes a chaotic state. Noticeably, MLEs $\lambda_b$ and $\lambda_r$ show that there is a transition from the periodic to the chaotic regime within the bifurcation range of $\gamma = 0.3809$ to $\gamma = 0.3821$, and we see that our models also exactly show the same region of transition. To ascertain this, we highlight the same transitioning region using the red dashed lines for all three plots, and they show excellent agreement with the transitions made by MLEs. However, the MLEs hint that there are a few intermediate periodic states around $\gamma=0.395$, which our models fail to capture.}
    \label{fig:duff_combined_result}
\end{figure}

Figure \ref{fig:duff_combined_result} presents the binary scores obtained from the three trained models -- LR, SVM, kNN -- plotted against the range of bifurcation values ($\gamma$). To validate our scores, we compare these scores with the maximum Lyapunov exponent (MLE), $\lambda_{max}$, that is computed using two methods. The first method is MLE using Benettin's method ($\lambda_{b}$) \cite{benettin1976kolmogorov} implemented in Python, which computes the Lyapunov exponent directly from the governing equations. This method allows for accurate tracking of transition regimes and yields precise exponent values for each bifurcation parameter due to access to full-state information and sufficiently long trajectories. The second method is the Rosenstein algorithm ($\lambda_{r}$) \cite{rosenstein1993practical} implemented using the NOLDS library \cite{scholzel_2019_3814723} in Python. Unlike the Benettin method, this approach estimates MLE directly from the time series data through Takens' reconstruction theorem \cite{takens2006detecting} without relying on the underlying equations. This makes $\lambda_{r}$ a suitable and realistic choice when only time series data are available. Therefore, we include it as a baseline for comparison against our proposed approach. Consequently, we compute $\lambda_{r}$ for the same set of time series used in the test set. This makes a more practical one-to-one comparison, as they are tested on the same time series signals. But, unlike in our classifier setup, we do not make the time series scanty here. We intentionally chose to retain the original time series length as it ensures higher fidelity in the MLE estimation, since random subsampling can introduce artifacts or unreliable estimates of the exponents.

Generally, in MLE, if $\lambda_{max}>0$, it corresponds to a chaotic state, and if $\lambda_{max}<0$ or $\lambda_{max} \approx 0$ then it corresponds to a periodic state. Our classifier models go a step further by assigning a binary label -- 1 for chaotic and 0 for periodic -- providing a sharp, interpretable separation between the two regimes. In Fig. \ref{fig:duff_combined_result}, both MLE estimations ($\lambda_{b}$ and $\lambda_{r}$) suggest that there is a transition in the range of $\gamma=0.3809$ to $\gamma=0.3821$ from the periodic to chaotic phase; additionally, there are also some intermediate periodic states near $\gamma=0.395$. To assess the consistency of our models with the MLEs, let us examine the binary scores of these models across the same bifurcation parameter space. Remarkably, our models show a clear demarcation between the periodic and chaotic regimes, with their transitions precisely matching with those made by the MLE estimations. To enhance clarity, we mark the transitioning region using the red dashed lines to highlight that all the models, along with MLEs, have the same parameter range of transition. This would help the readers visually compare and ascertain that the transitions made by our models coincide with the MLE-based transitions. 

We would also like to report that our models failed to identify or pick all the intermediate periodic states around $\gamma=0.395$. These windows occupy only a small or narrow portion of the bifurcation interval and consist of relatively few sample states. As a result, their statistical representation in the training data is limited. Consequently, our classifiers are primarily trained to distinguish the broader periodic and chaotic regimes, and hence, they may not be sufficiently exposed to these subtle transitions during the training, causing them to occasionally misclassify. We would like to insist that the primary aim of this study is to identify the dynamic state change demarcating the chaotic regime from the periodic regime, which our models perfectly do. We consider finding these small intermediate periodic orbits as an additional advantage if detected, and not as our primary focus. However, consistent identification of these small intermediate periodic windows requires a denser sampling of those parameter regions. We would like to report that kNN exhibits a few misclassifications in both periodic and chaotic regimes. With all these explanations laid out, we move on to the other systems, sticking to the same scheme of analysis.  

\begin{table}
\centering
\begin{tabular}{|l|lll|}
\hline  
\multirow{2}{*}{\textbf{Split (K=5)}} & \multicolumn{3}{l|}{\textbf{\begin{tabular}[c]{@{}l@{}}Accuracy $\%$ \end{tabular}}} \\ \cline{2-4}
                                      & \multicolumn{1}{l|}{\textbf{LR}}   & \multicolumn{1}{l|}{\textbf{SVM}} & \textbf{kNN}  \\ \hline
\textbf{Split 1}                      & \multicolumn{1}{l|}{100}          & \multicolumn{1}{l|}{100}           & 90          \\ \hline
\textbf{Split 2}                      & \multicolumn{1}{l|}{100}          & \multicolumn{1}{l|}{100}           & 95          \\ \hline
\textbf{Split 3}                      & \multicolumn{1}{l|}{97.5}            & \multicolumn{1}{l|}{95}           & 85          \\ \hline
\textbf{Split 4}                      & \multicolumn{1}{l|}{97.5}          & \multicolumn{1}{l|}{97.5}           & 90            \\ \hline
\textbf{Split 5}                      & \multicolumn{1}{l|}{100}           & \multicolumn{1}{l|}{100}          & 92.5          \\ \hline
\textbf{Average}                      & \multicolumn{1}{l|}{\textbf{99}} & \multicolumn{1}{l|}{\textbf{98.5}}  & \textbf{90.5} \\ \hline
\end{tabular}
\caption{We report the accuracy score for each split in 5-fold cross-validation applied for three classifiers: LR, SVM, and kNN on the Duffing system. The accuracy scores obtained for 5 trials are consistent for all three classifiers: LR, SVM, and kNN. The high average accuracy scores obtained, exceeding $90\%$, ascertain and complement the choice of hyperparameter.}
\label{tab:table_duff_acc}
\end{table}

\subsection{Rossler system}
\label{results_ross}
\begin{figure}[hbt!]
     \centering
     \begin{subfigure}[b]{0.45\textwidth}
         \centering
         \includegraphics[width=\textwidth]{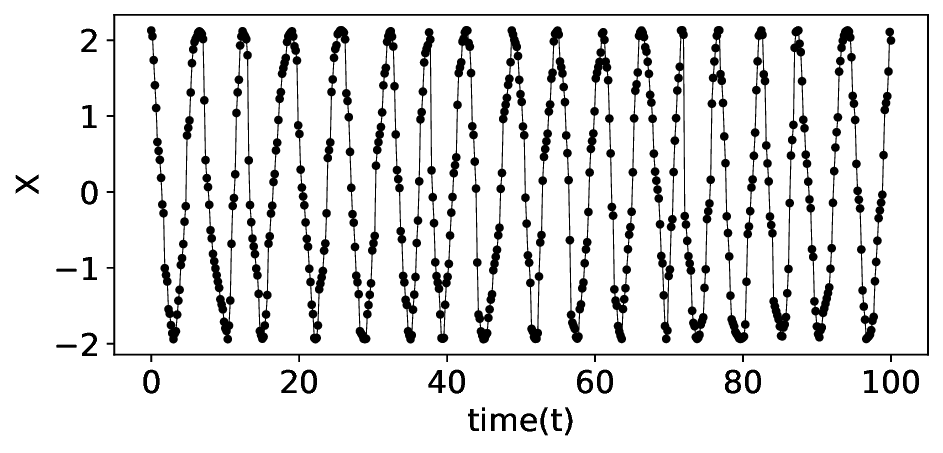}
         \caption{}
         \label{fig:ross_1}
     \end{subfigure}
	 \begin{subfigure}[b]{0.45\textwidth}
         \centering
         \includegraphics[width=\textwidth]{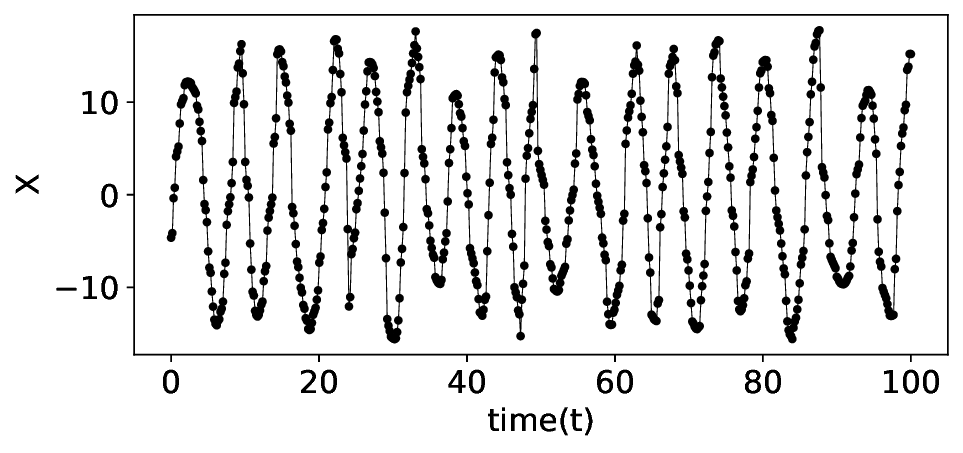}
         \caption{}
         \label{fig:ross_11}
     \end{subfigure}
     	        \caption{Scanty time series data of the Rossler system exhibiting periodic behavior at $C=1$ and chaotic behavior at $C=11$.}
        		\label{fig:ross_bifurcation}
     \end{figure}

In 1976, Otto Rossler \cite{rossler1976equation} devised a three-dimensional non-linear system called the Rossler system, which was analogous to the Lorenz model, to facilitate quantitative analysis. The Rossler system demonstrates both periodic and chaotic behavior for different values of its bifurcation parameters. The governing equations of the system are:
$$\dot{x} = -y - z $$
$$\dot{y} = x + ay $$
$$\dot{z} = b + z(x - c)  $$
The bifurcation route to chaos occurs when the bifurcation parameter, $c$ is varied and other parameters are kept constant at $a=0.1$, and $b=0.1$. The initial conditions for the system are $x(0) = 1.0$, $y(0) = 1.0$, $z(0) = 1.0$. The system is evolved over a time range of 0 to 1000s with a step size of 0.05, yielding a total of 20000 data points (time series). We omit the first 18000 points as transient data and consider the last 2000 data points (which is the last 100s window) as the representative signal. To mimic the real-world scenario, we remove $70\%$ of the data (from the 2000 data points), which accounts for 1400 data points, leaving scanty or sparse time series data of 600 points. These 600 data points represent a real-world incomplete time series or time series with missing elements. Rossler shows dynamic behavior for different bifurcation parameters. The system exhibits a wide range of periodic regimes in the form of period 1 for $c=1$, period 2 for $c=6$. Beyond $c=8.8$, the system exhibits a chaotic phase (with intermittent periodic states in between). However, for illustration purposes, we show a periodic scanty time series signal at $c=1$ and a chaotic scanty time series signal at $c=11$ in Fig. \ref{fig:ross_bifurcation}.

We compute the sublevel set persistence for each scanty time series obtained over a range of bifurcation parameters from $c=1$ to $c=11.75$ in steps of $0.05$, which accounts for 216 parameters in total. The obtained topological features from the persistence diagrams are vectorized into 5 input features, and they are fed into the classifiers as training and test sets. We employ the same strategy mentioned in \ref{duff_results_section} to create the CSV file (master dataset). The target class is manually labeled with tag `0' if it is periodic and tag `1' if it is chaotic. We use $80\%$ of the master dataset as the training set and use $20\%$ of the master dataset (which accounts for 44 test data) as the test set to evaluate the model. We use three different classifiers of LR, SVM, and kNN to perform the classification. We maintain a constant random seed for the master dataset split to ensure that all three models are tested and evaluated on the same grounds.  

\begin{figure}[hbt!]
\begin{minipage}{.5\linewidth}
\centering
\subfloat[Confusion matrix for LR]{\label{fig:cnf_LR_ross}\includegraphics[scale=.22]{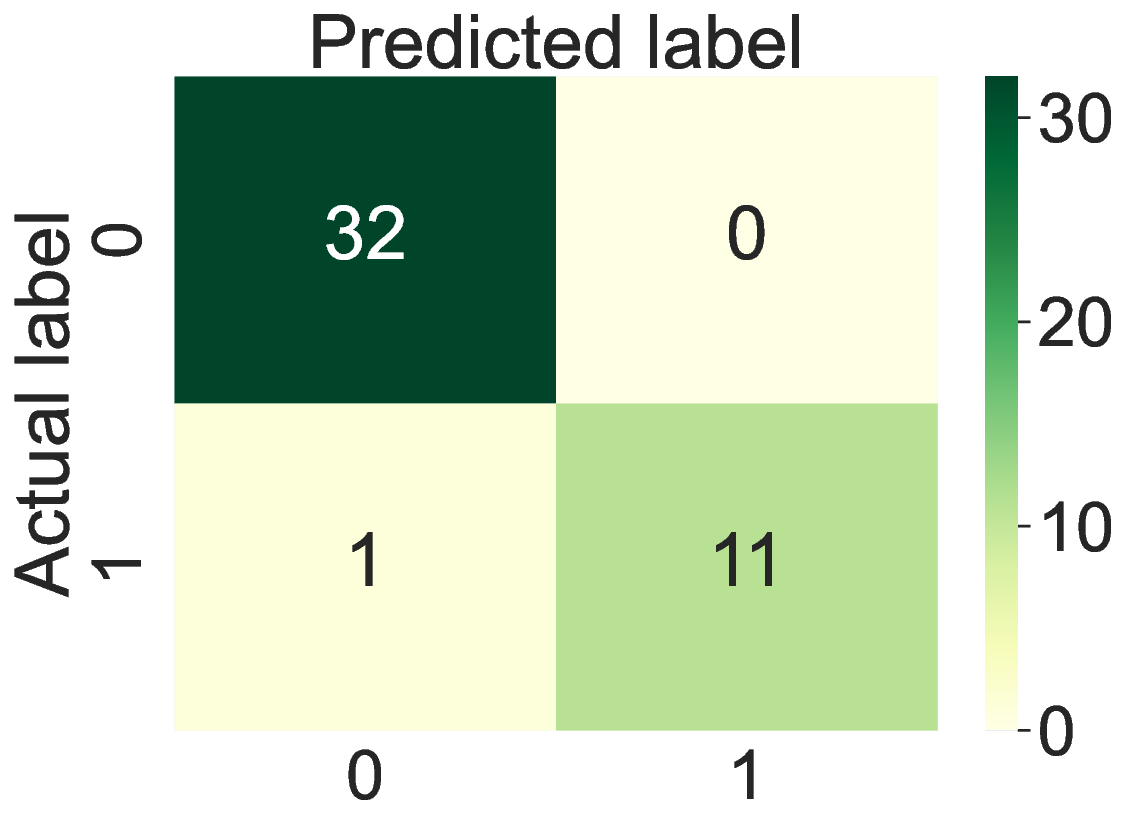}}
\end{minipage}%
\begin{minipage}{.5\linewidth}
\centering
\subfloat[Confusion matrix for SVM]{\label{fig:cnf_SVM_ross}\includegraphics[scale=.22]{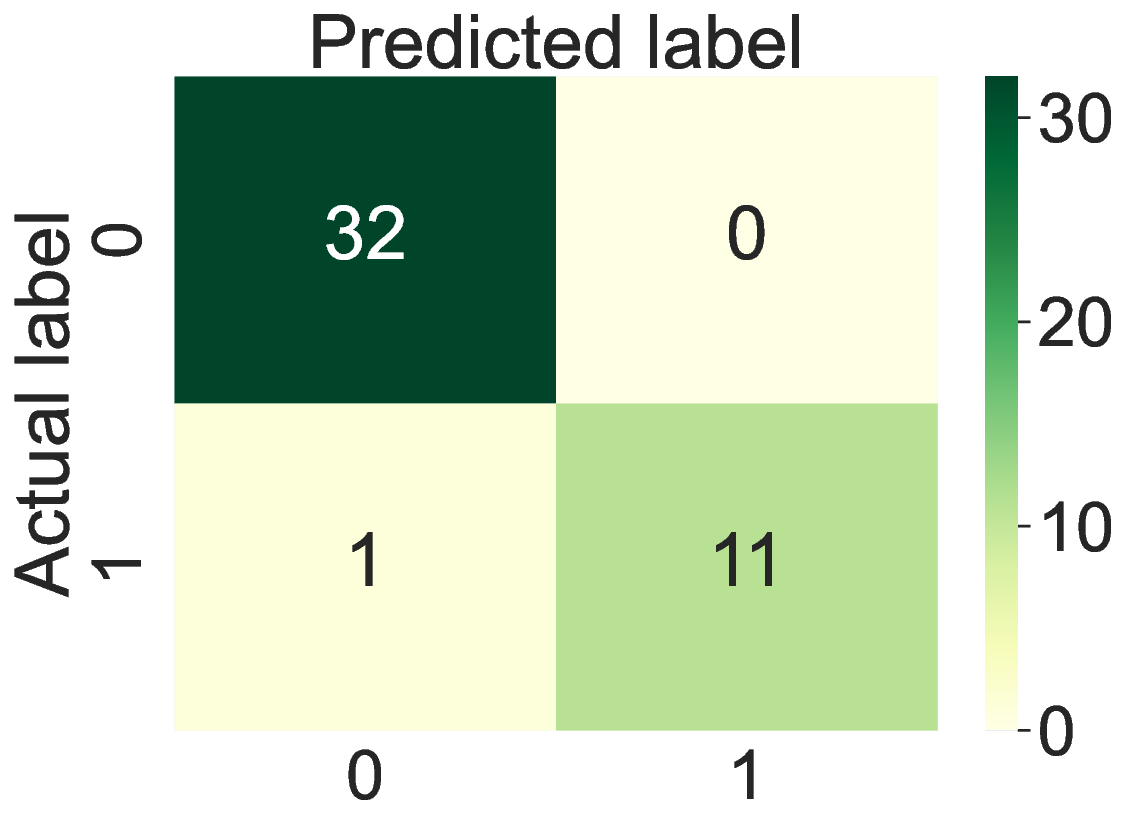}}
\end{minipage}\par\medskip
\centering
\subfloat[Confusion matrix for kNN]{\label{cnf_kNN_ross}\includegraphics[scale=.22]{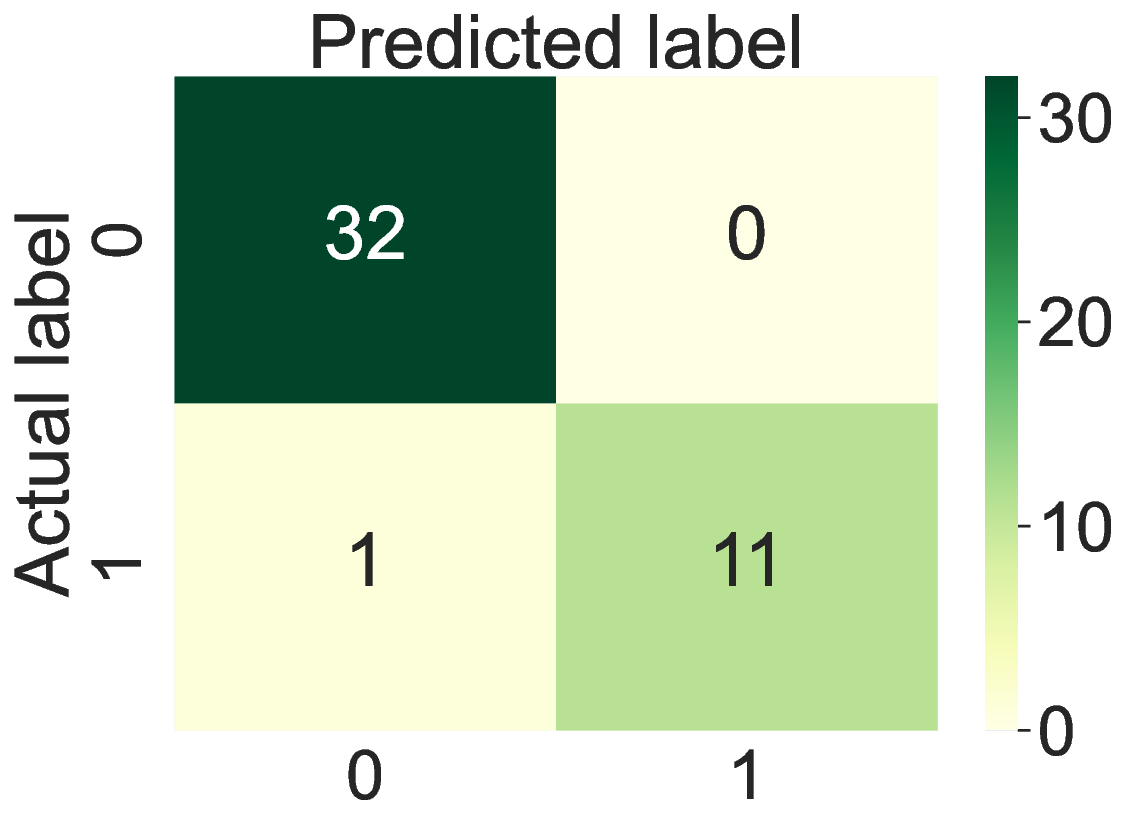}}
\caption{This figure compares the confusion matrix of LR, SVM, and kNN for the Rossler system obtained by evaluating the same test set data: We choose a test size of $20\%$ from the total dataset, which accounts for 44 data samples. They all show excellent classification capabilities, where all three classifiers made 32 correct predictions as periodic samples and 11 correct predictions as chaotic samples. However, all three classifiers misclassified 1 chaotic sample as periodic.}
\label{fig:cnf_mat_ross}
\end{figure}

\begin{figure}[hbt!]
\centering
\includegraphics[width=0.45\textwidth]{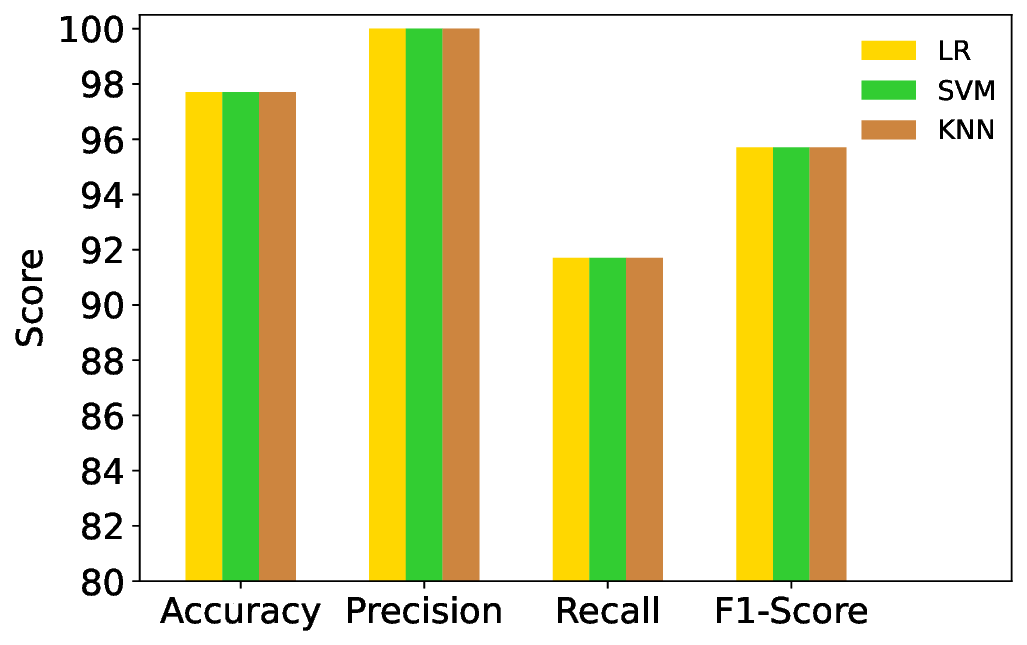}
    \caption{We present the evaluation metric scores of LR, SVM, and kNN for the Rossler system. They all obtained high scores, and they turned up to have the same scores as well.}
    \label{fig:metrics_ross}
\end{figure}

We present the summary of the confusion matrix for all three classifiers that were obtained by evaluating on the test set in Fig. \ref{fig:cnf_mat_ross} to showcase their correctness in predicting the respective categories. All three classifiers correctly predicted the 32 samples as periodic and the 11 samples as chaotic. On the other hand, we notice that all three classifiers have wrongly predicted 0 samples as chaotic and 1 sample as periodic. We then move on to the evaluation metrics: accuracy, precision, recall, and F1-score, where we witness the performance of all three classifiers compared using the bar graph in Fig. \ref{fig:metrics_ross}. The plots show excellent scores for all three classifiers; surprisingly, they all have turned out to produce the same metric scores. The LR classifier was run on default parameters, while kNN and SVM hyperparameters were tuned to achieve the best accuracy scores. Additionally, the performance of the classifiers was evaluated over 30 independent random seeds, and the average accuracy obtained for LR is $95.98\% \pm 2.54\%$, SVM is $96.21\% \pm 2.37\%$, and for kNN is $96.21\% \pm 2.37\%$. The hyperparameter details are provided in Table \ref{tab:hyp_param_ross} for the practitioners to reproduce the results. We use the K-fold cross-validation to ensure that the hyperparameters chosen consistently produce consistent accuracy scores for different validation sets. We choose $K=5$ and state their respective accuracy scores for each fold using three classifiers in Table \ref{tab:table_acc_ross}. We achieved an average accuracy score of $90.7\%$ for LR, $96.2\%$ for SVM, and $95.9\%$ for kNN. The obtained high accuracy and consistent scores for each trial certify that the chosen hyperparameter generalizes well for the validation set data.

\begin{table*}
\centering
\begin{tabular}{|l|ll|ll|}
\hline
\textbf{LR}                         & \multicolumn{2}{l|}{\textbf{SVM}}                                       & \multicolumn{2}{l|}{\textbf{kNN}}                                    \\ \hline
\multirow{4}{*}{Default parameters} & \multicolumn{1}{l|}{\textbf{Hyper-parameter(s)}} & \textbf{Value(s)}    & \multicolumn{1}{l|}{\textbf{Hyper-parameter(s)}} & \textbf{Value(s)} \\ \cline{2-5} 
                                    & \multicolumn{1}{l|}{Kernel   function}           & rbf                  & \multicolumn{1}{l|}{No. of   neighbors (k)}      & 7                 \\ \cline{2-5} 
                                    & \multicolumn{1}{l|}{\multirow{2}{*}{C}}          & \multirow{2}{*}{100} & \multicolumn{1}{l|}{Weights}                     & uniform           \\ \cline{4-5} 
                                    & \multicolumn{1}{l|}{}                            &                      & \multicolumn{1}{l|}{Distance   metric}           & Euclidean         \\ \hline
\end{tabular}
\caption{Table representing hyperparameter details for LR, SVM, and kNN for the Rossler system.}
\label{tab:hyp_param_ross}
\end{table*}

\begin{table}
\centering
\begin{tabular}{|l|lll|}
\hline
\multirow{2}{*}{\textbf{Split (K=5)}} & \multicolumn{3}{l|}{\textbf{\begin{tabular}[c]{@{}l@{}}Accuracy $\%$ \end{tabular}}} \\ \cline{2-4} 
                                      & \multicolumn{1}{l|}{\textbf{LR}}          & \multicolumn{1}{l|}{\textbf{SVM}}         & \textbf{kNN}        \\ \hline
\textbf{Split 1}                      & \multicolumn{1}{l|}{97.1}                 & \multicolumn{1}{l|}{97.7}                 & 97.1                \\ \hline
\textbf{Split 2}                      & \multicolumn{1}{l|}{77.1}                  & \multicolumn{1}{l|}{95.3}                 & 100                 \\ \hline
\textbf{Split 3}                      & \multicolumn{1}{l|}{91.1}                 & \multicolumn{1}{l|}{97.6}                 & 94.1                \\ \hline
\textbf{Split 4}                      & \multicolumn{1}{l|}{94.1}                 & \multicolumn{1}{l|}{93}                   & 94.1                \\ \hline
\textbf{Split 5}                      & \multicolumn{1}{l|}{94.1}                 & \multicolumn{1}{l|}{97.6}                 & 94.1                \\ \hline
\textbf{Average}                      & \multicolumn{1}{l|}{\textbf{90.7}}        & \multicolumn{1}{l|}{\textbf{96.2}}        & \textbf{95.9}       \\ \hline
\end{tabular}
\caption{We report the accuracy score for each split in 5-fold cross-validation applied for three classifiers: LR, SVM, and kNN on the Rossler system. The accuracy scores obtained for 5 trials are consistent, and the high average accuracy scores ascertain the hyperparameter choices and indicate that the model generalizes well to the validation data.}
\label{tab:table_acc_ross}
\end{table}

\begin{figure}[hbt!]
\centering
\includegraphics[width=0.5\textwidth]{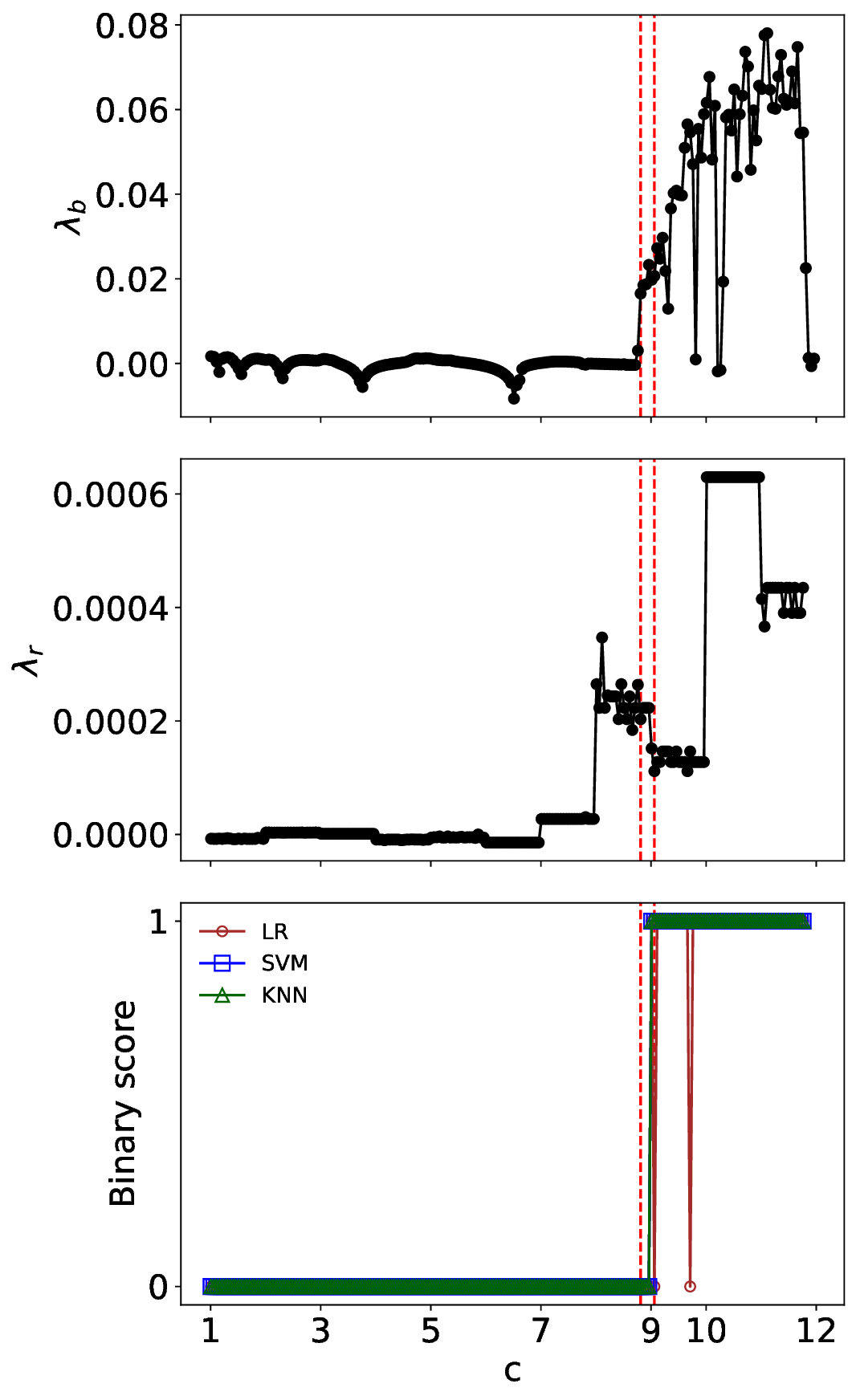}
    \caption{We showcase the classification made by the trained classifiers of LR, SVM, and kNN using the binary scores for \emph{new test parameters} of the Rossler system, where 0 connotes a periodic state, while 1 connotes a chaotic state. Noticeably, MLEs $\lambda_b$ and $\lambda_r$ show that there is a transition from the periodic to the chaotic regime within the bifurcation range of $c = 8.81$ to $c = 9.06$, and we see that our models also exactly show the same region of transition. To ascertain this, we highlight the same transitioning region using the red dashed lines for all three plots, and they show excellent agreement with the transitions made by MLEs. However, $\lambda_b$ hints that there are a few intermediate periodic states around $c=10$, which our models fail to capture.}
    \label{fig:ross_combined_result}
\end{figure}

Using these trained LR, SVM, and kNN models, we extend the chaotic transition prediction for a non-overlapping set of bifurcation parameters. We consider scanty time series from bifurcation parameters ranging from $c=1.01$ to $c=11.76$ in steps of 0.02 that account for 216 parameters. We then compute the sublevel set homology for all these time series and feature extract the topological features (birth-death coordinates), after which they are vectorized into the 5 input features. These input features are then sent into our models for classification. Now, we compare the ML models' binary scores with the maximal Lyapunov exponent (MLE) computed with two different methods: $\lambda_r$ and $\lambda_b$. Figure \ref{fig:ross_combined_result} shows the ML classified Binary scores using different classifiers along with the $\lambda_r$ and $\lambda_b$ values, plotted against the bifurcation parameter ($c$). The plot of $\lambda_b$ shows that there is a transition from periodic to the chaotic regime around the range $c=8.81$ to $c=9.06$. However, a quick look at the $\lambda_r$ plot, one might be tempted to say that there is a transition that commences from $c=7$ and then shows higher jumps as we move along $c=8$.

This plot will force users to mark the window: $c=7$ to  $c=8$ as the region that demarcates the chaotic regime from the periodic, which is misleading and inaccurate. Consequently, we would like to turn our attention towards the classifiers to see how they perform. It takes no expertise to say that all three classifiers have made transitions towards the chaotic regime that mirror the transition made by $\lambda_b$. In addition to this, we highlight the transitioning region using the red dashed lines to ascertain that all the plots have the same parametric transition. Now, this is a classic case where a traditional method, the maximum Lyapunov exponent calculated using Rosenstein's method ($\lambda_r$), can falter and where our method can show fidelity in replicating the dynamic nature of the system. We would also like to report that our models were not able to identify the intermediate periodic states (three or four states) in the chaotic region; however, as said earlier, we aim to differentiate or alarm when there is a transition to the chaotic regime as our primary focus.

\subsection{Lorenz system}
\label{results_lorenz}
\begin{figure}[hbt!]
     \centering
     \begin{subfigure}[b]{0.45\textwidth}
         \centering
         \includegraphics[width=\textwidth]{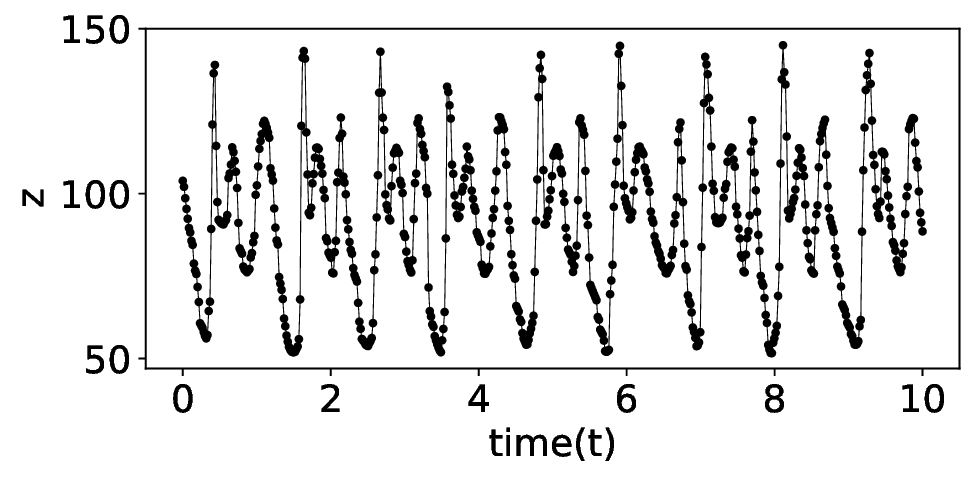}
         \caption{}
         \label{fig:rho_996}
     \end{subfigure}
	 \begin{subfigure}[b]{0.45\textwidth}
         \centering
         \includegraphics[width=\textwidth]{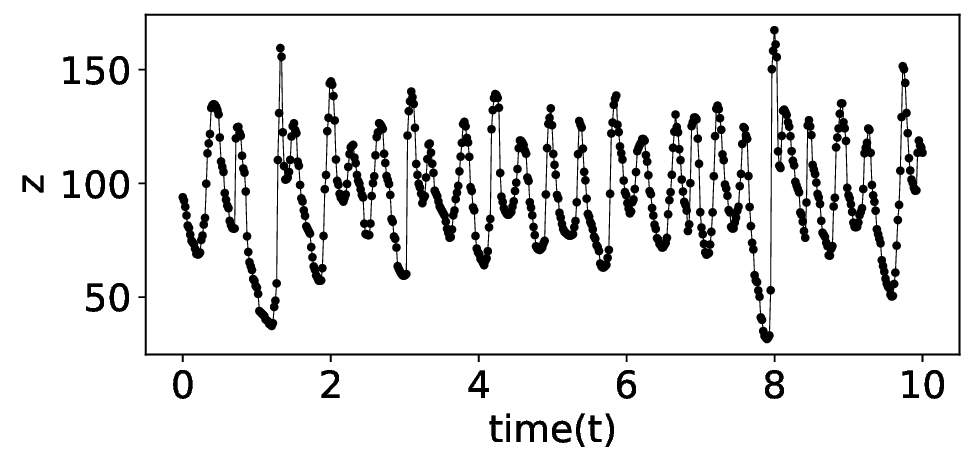}
         \caption{}
         \label{fig:rho_102}
     \end{subfigure}
     	        \caption{Scanty time series data of the Lorenz system exhibiting periodic behavior at  $\rho=99.6$ and chaotic behavior at $\rho=102$.}
        		\label{fig:lor_bifurcation}
     \end{figure}

In 1963, Edward Lorenz reduced a 12-dimensional system used to model atmospheric convection to three ordinary differential equations, known as the Lorenz equations. It is a three-dimensional, nonlinear, aperiodic, deterministic system that exhibits periodic and chaotic phases for different bifurcation parameters. The governing equations are given by:
$$\dot{x} = \sigma (y - x)  $$
$$\dot{y} = x(\rho - z) - y $$
$$\dot{z} = xy - \beta z  $$
The bifurcation route to chaos occurs by changing the bifurcation parameter, rho ($\rho$), and keeping the other parameters constant at $\sigma =10$, $\beta = 8/3$, $x(0) = 1$, $y(0) = 1$, $z(0) = 1$. We evolve the system over a time range of $t=0$ to $t=100$ over a step size of 0.005, which results in a time series of 20000 data points. We ignore the first 18000 points as transient data and consider the last 2000 time series data (which is nothing but the last $10s$ window) as the representative signal. We create scanty data from the representative signal, which is 2000 data points, by randomly removing $70\%$ data, which accounts for 1400 data points. This results in 600 data points of time series data that mimics real-world scanty data or data with missing elements. We show the system's scanty time series data for both periodic and chaotic behavior at different bifurcation parameters in Fig. \ref{fig:lor_bifurcation}.

We compute the sublevel set homology for each scanty time series for the bifurcation parameters ranging from $\rho=99.55$ to $\rho=102$ in steps of 0.01, which accounts for 246 parameters in total. The obtained topological features in the PDs are vectorized into 5 input features and then fed into the model using the same method mentioned in Sec.\ref{vectorization} and Sec. \ref{duff_results_section}, categorizing the periodic features with target label `0' and the chaotic features with label `1'. We use the $80\%-20\%$ train-test split for LR, SVM, and kNN classifiers, which accounts for 50 test data samples. As stated earlier, we maintain a constant random seed for the master dataset split for all three models to maintain uniformity. 

\begin{figure}
\begin{minipage}{.5\linewidth}
\centering
\subfloat[Confusion matrix for LR]{\label{fig:cnf_LR_lor}\includegraphics[scale=.22]{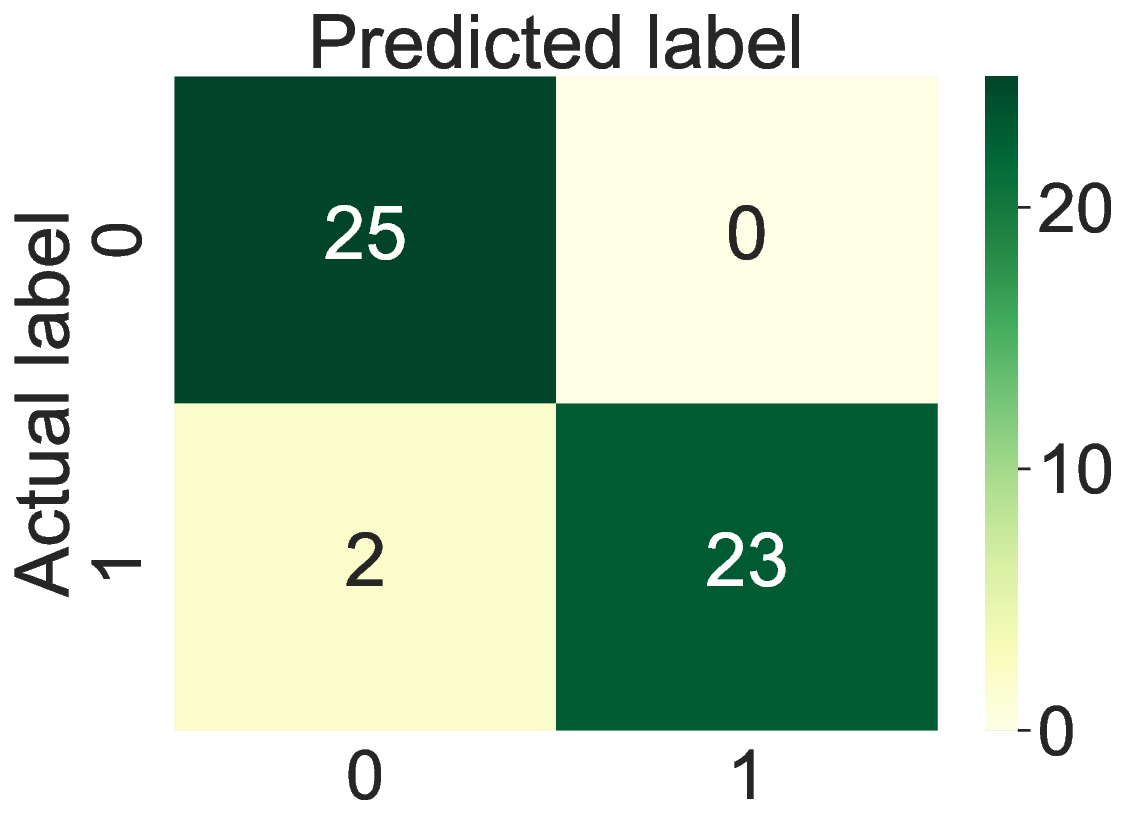}}
\end{minipage}%
\begin{minipage}{.5\linewidth}
\centering
\subfloat[Confusion matrix for SVM]{\label{fig:cnf_SVM_lor}\includegraphics[scale=.22]{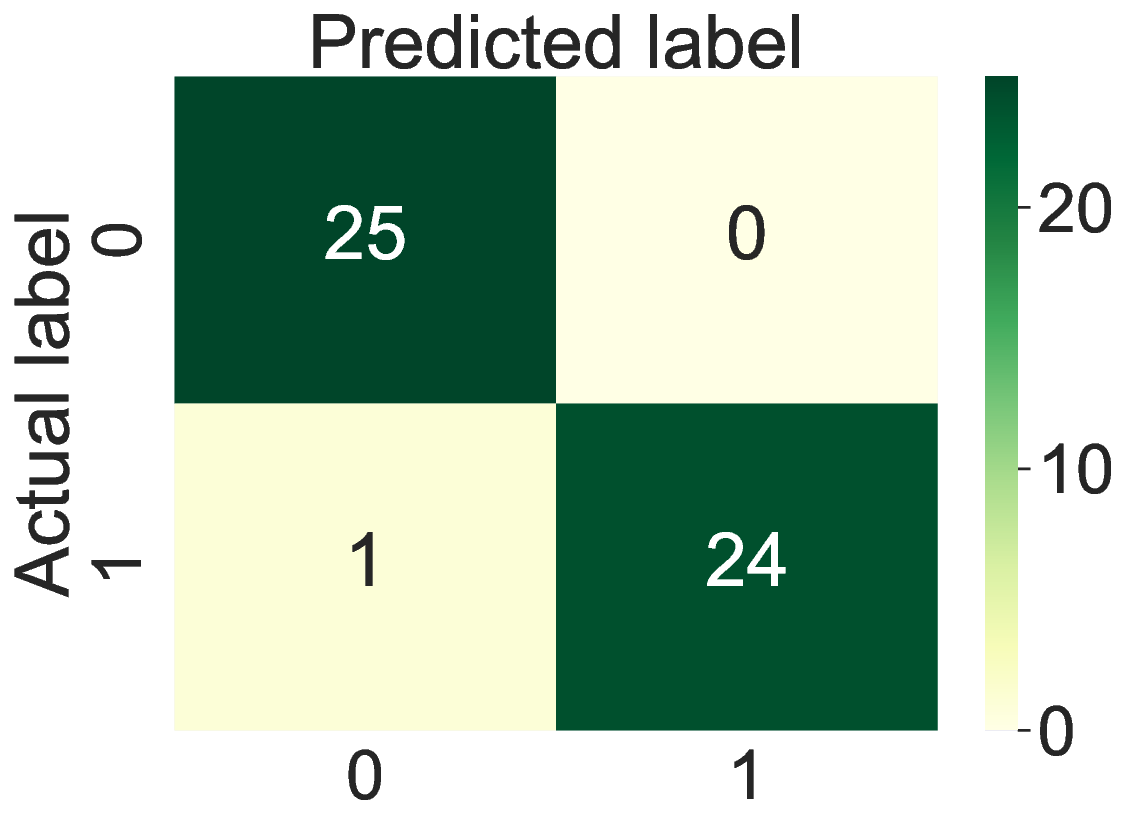}}
\end{minipage}\par\medskip
\centering
\subfloat[Confusion matrix for kNN]{\label{cnf_kNN_lor}\includegraphics[scale=.22]{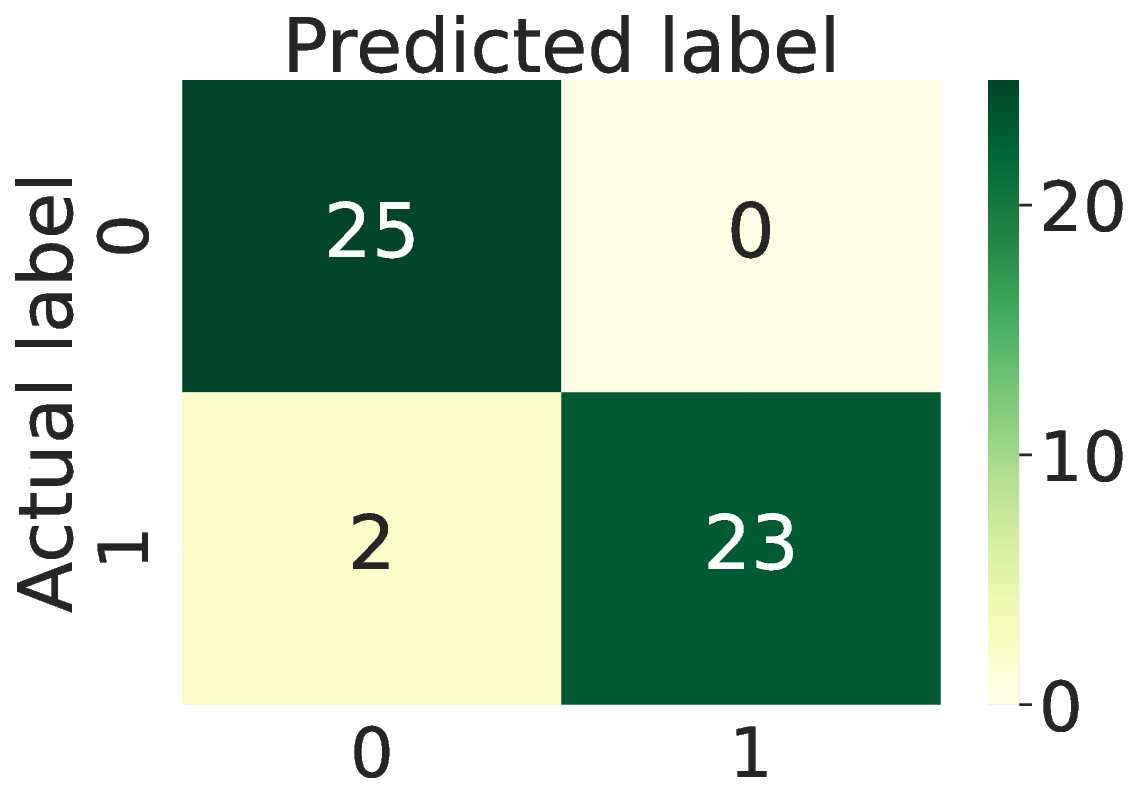}}
\caption{This figure compares the confusion matrix of LR, SVM, and kNN for the Lorenz system obtained by evaluating the same test set data: We choose a test size of $20\%$ from the total dataset, which accounts for 50 data samples. All three classifiers made 25 correct predictions as periodic samples and 24 as chaotic samples. However, all three classifiers misclassified 1 chaotic sample as periodic.}
\label{fig:cnf_mat_lor}
\end{figure}

\begin{figure}
\centering
\includegraphics[width=0.42\textwidth]{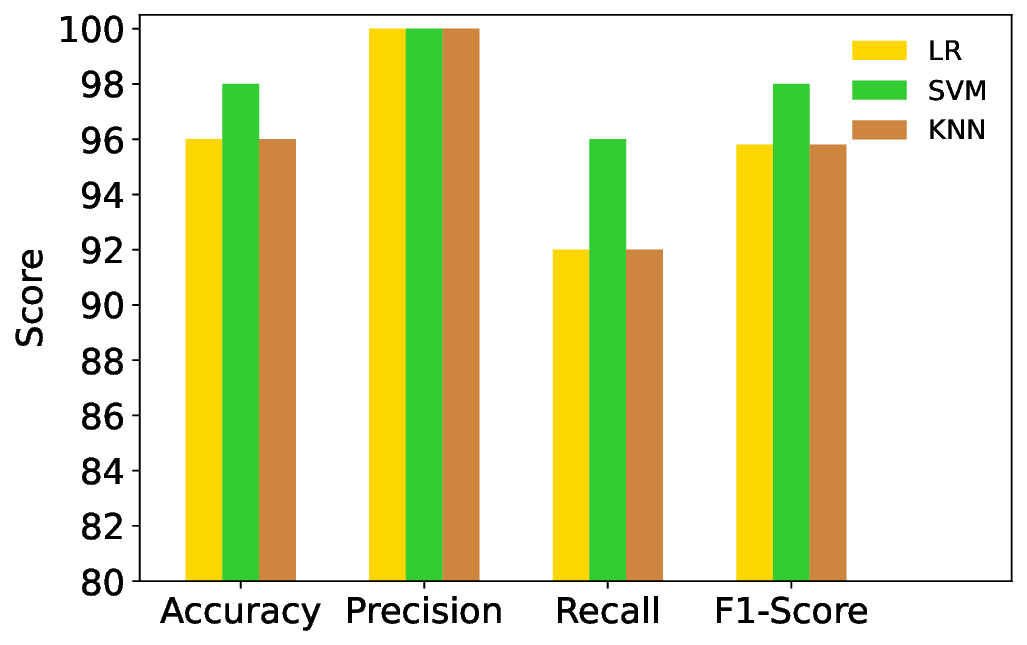}
    \caption{We present the evaluation metric scores of LR, SVM, and kNN for the Lorenz system. They all obtained excellent scores, showcasing their performance, with SVM displaying standout scores in three departments.}
    \label{fig:metrics_lor}
\end{figure}

The confusion matrix is shown in Fig. \ref{fig:cnf_mat_lor} to showcase the correctness in predicting their respective categories. All three classifiers correctly predicted 25 samples as periodic. For chaotic samples, both LR and KNN correctly predicted 23, while SVM correctly predicted 24 samples. However, all three wrongly predicted 0 samples as chaotic, while both LR and kNN misclassified 2 samples, and SVM misclassified 1 sample as periodic. We plot the metric scores of accuracy, precision, recall, and F1-score in Fig. \ref{fig:metrics_lor} of the three classifiers for comparison. All classifiers achieved high scores across all metrics, with SVM dominating its counterparts in three departments. Additionally, we tested the performance of the classifiers over 30 independent random seeds, and the average accuracy obtained for LR is $98.2\% \pm 1.66\%$, SVM is $99\% \pm 1.34\%$, and for kNN is $98.47\% \pm 1.43\%$. The LR model was run on the default parameter settings, while the kNN and SVM hyperparameters were tuned to obtain such high metric scores. We present the tuned hyperparameter details in Table \ref{tabl:hyp_par_lor} to facilitate reproducibility for readers. We present K-fold cross-validation scores to ensure that the hyperparameters chosen generalize well and consistently produce high scores for different validation sets. We tabulate the accuracy scores for each trial of the K-fold technique in Table \ref{tab:table_acc_lor} along with its average accuracy score of: $98.5\%$ for LR, $91.5\%$ for SVM, and $99\%$ for kNN. The obtained high average scores demonstrate the effectiveness of hyperparameter choices and the model's ability to adapt to unseen data.

\begin{table*}
\centering
\begin{tabular}{|l|ll|ll|}
\hline
\textbf{LR}                         & \multicolumn{2}{l|}{\textbf{SVM}}                                       & \multicolumn{2}{l|}{\textbf{kNN}}                                    \\ \hline
\multirow{4}{*}{Default parameters} & \multicolumn{1}{l|}{\textbf{Hyper-parameter(s)}} & \textbf{Value(s)}    & \multicolumn{1}{l|}{\textbf{Hyper-parameter(s)}} & \textbf{Value(s)} \\ \cline{2-5} 
                                    & \multicolumn{1}{l|}{Kernel   function}           & rbf                  & \multicolumn{1}{l|}{No. of   neighbors (k)}      & 13                 \\ \cline{2-5} 
                                    & \multicolumn{1}{l|}{\multirow{2}{*}{C}}          & \multirow{2}{*}{100} & \multicolumn{1}{l|}{Weights}                     & distance          \\ \cline{4-5} 
                                    & \multicolumn{1}{l|}{}                            &                      & \multicolumn{1}{l|}{Distance   metric}           & euclidean         \\ \hline
\end{tabular}
\caption{Table representing hyperparameter details for LR, SVM, and kNN for Lorenz system.}
\label{tabl:hyp_par_lor}
\end{table*}

\begin{table}
\centering
\begin{tabular}{|l|lll|}
\hline
\multirow{2}{*}{\textbf{Split (K=5)}} & \multicolumn{3}{l|}{\textbf{Accuracy $\%$}}                                         \\ \cline{2-4} 
                                      & \multicolumn{1}{l|}{\textbf{LR}} & \multicolumn{1}{l|}{\textbf{SVM}}  & \textbf{kNN} \\ \hline
\textbf{Split 1}                      & \multicolumn{1}{l|}{100}         & \multicolumn{1}{l|}{88}            & 100          \\ \hline
\textbf{Split 2}                      & \multicolumn{1}{l|}{100}         & \multicolumn{1}{l|}{95.9}           & 100          \\ \hline
\textbf{Split 3}                      & \multicolumn{1}{l|}{97.4}         & \multicolumn{1}{l|}{91.8}          & 100          \\ \hline
\textbf{Split 4}                      & \multicolumn{1}{l|}{97.4}        & \multicolumn{1}{l|}{87.8}           & 97.4         \\ \hline
\textbf{Split 5}                      & \multicolumn{1}{l|}{97.4}        & \multicolumn{1}{l|}{93.9}            & 97.4         \\ \hline
\textbf{Average}                      & \multicolumn{1}{l|}{\textbf{98.5}} & \multicolumn{1}{l|}{\textbf{91.5}} & \textbf{99}  \\ \hline
\end{tabular}
\caption{We report the accuracy score for each split in 5-fold cross-validation applied for three classifiers: LR, SVM, and kNN on the Lorenz system. The accuracy scores obtained for 5 trials are consistent, and we obtained excellent average accuracy scores exceeding $90$, ascertaining the hyperparameter choices and indicating that the model generalizes well to validation data.}
\label{tab:table_acc_lor}
\end{table}

\begin{figure}[hbt!]
\centering
\includegraphics[width=0.5\textwidth]{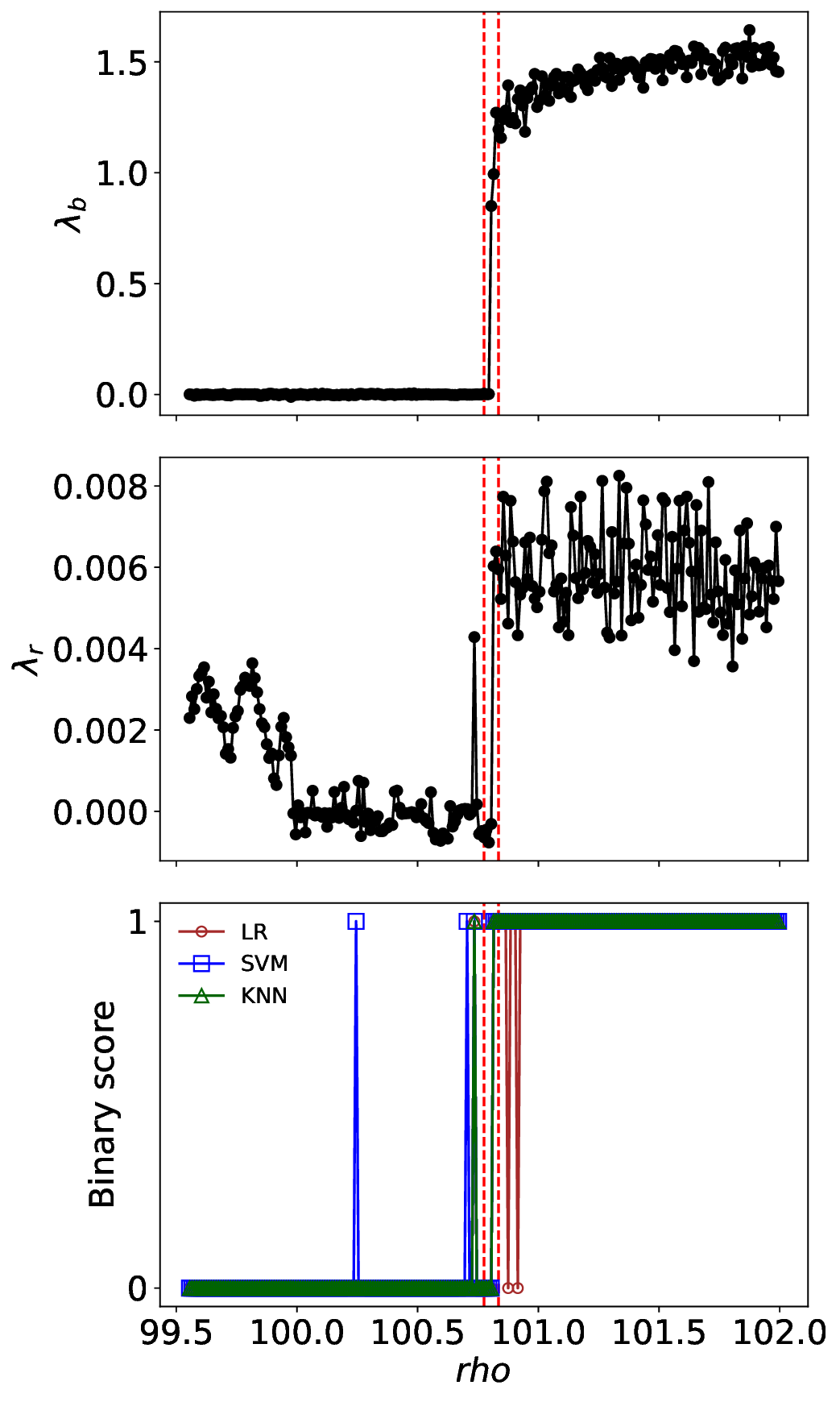}
    \caption{We showcase the classification made by the trained classifiers of LR, SVM, and kNN using the binary scores for \emph{new test parameters} of the Lorenz system, where 0 connotes a periodic state while 1 connotes a chaotic state. Noticeably, MLEs $\lambda_b$ and $\lambda_r$ show that there is a transition from the periodic to the chaotic regime within the bifurcation range of $\rho=100.775$ to $\rho=100.835$, and we see that our models also exactly show the same region of transition. To ascertain this, we highlight the same transitioning region using the red dashed lines for all three plots, and they show excellent agreement with the transitions made by MLEs.}
    \label{fig:lor_combined_result}
\end{figure}

Using the trained LR, SVM, and kNN models, we extend the chaotic prediction for different dynamical states to other resolutions of bifurcation parameters. We intentionally use a finer resolution of bifurcation parameters by considering the intermediate values not used in the earlier train-test dataset to showcase the predictability and versatility of the model. We take a scanty time series of bifurcation parameters from $\rho=99.555$ to $\rho=101.995$ in steps of 0.01, which accounts for 245 parameters, and we compute the sublevel set homology for the time series. The corresponding topological features that were obtained are sent into our models for classification. Figure \ref{fig:lor_combined_result} shows the binary score for all three classifiers, with 0 denoting a periodic nature and 1 denoting a chaotic nature for each bifurcation parameter. The binary scores of these classifiers are matched against the two MLEs: $\lambda_b$ and $\lambda_r$. The plot of $\lambda_b$ shows that there is a transition from periodic to chaotic phase around the region of $\rho=100.775$ to $\rho=100.835$. However, when we have a quick look at the $\lambda_r$ plot, one might be tempted to say that the transitions seem to resonate with $\lambda_b$ but with different levels of magnitude. Consequently, a careful and closer examination of the values would hint to the observers that $\lambda_r$ suggests that there are chaotic states in the range of $\rho=99.555$ to $\rho=100$, which is inaccurate. We say this because the general notion of MLE is that, when $\lambda_{max}>0$, it denotes chaotic behavior. Hence, in our case, we see the $\lambda_r$ being positive for the range of values between $\rho=99.555$ to $\rho=100$. This misleads the users because $\lambda_b$ clearly shows no signs of such chaotic behavior in that region. Now, let us turn our attention towards our approach, where our models show a transition towards the chaotic regime in the window of $\rho$ values ranging from $\rho=100.775$ to $\rho=100.835$, which exactly matches the transition shown by $\lambda_b$. Furthermore, we mark this window using red-dashed lines to highlight that the models and $\lambda_b$ have the same parametric transition. This is another instance where our approach has proven that it has an edge over $\lambda_r$, showcasing its potential as a viable alternative for the research community. We also admit that we do have some misclassifications (three or four samples) before and after the transitioning region, with kNN showing the least number of misclassifications (just one). Now, let us move on to the last part of this section, where we apply our approach on real-world ECG data to see how it performs.
     
\subsection{ECG Heartbeat Classification}
An electrocardiogram (ECG) is used to study and monitor the function of the cardiovascular system \cite{kligfield2007recommendations}. This essentially measures the electrical activity of the heart as a function of time, thereby generating time series data. Cardiologists typically analyze the PQRST wave produced by ECG data to detect abnormalities or irregularities in the heartbeat. However, manual validation and evaluation of these data are highly time-consuming, and they are prone to human errors. This demands an automated machine-learning approach to classify the anomalies, enhancing diagnostic efficiency and accuracy.

In this study, we evaluate the effectiveness of our proposed method in classifying the normal and abnormal heartbeats using PTB Diagnostics ECG datasets \cite{ptb_data,physiobank}. We used the preprocessed version of this dataset, which is made available on Kaggle \cite{kaggle_ecg}. The original PTB database contains ECG records from 268 subjects, which includes 148 cases diagnosed with myocardial infarction, 68 cases diagnosed with 7 different heart diseases (which collectively sum up to 216 unhealthy subjects), and 52 healthy subjects. The Kaggle version provides a standardized, preprocessed form that involves cropping, downsampling, and padding with zeros to a fixed dimension (column) of 188. Each sample in the Kaggle dataset corresponds to a single PQRST complex labeled as either normal or abnormal (grouping all the above-mentioned heart diseases into the abnormal class). We used a balanced data set that contains 2,000 normal samples and 2,000 abnormal data samples for our work. 

\begin{figure}
     \centering
     \begin{subfigure}[b]{0.45\textwidth}
         \centering
         \includegraphics[width=\textwidth]{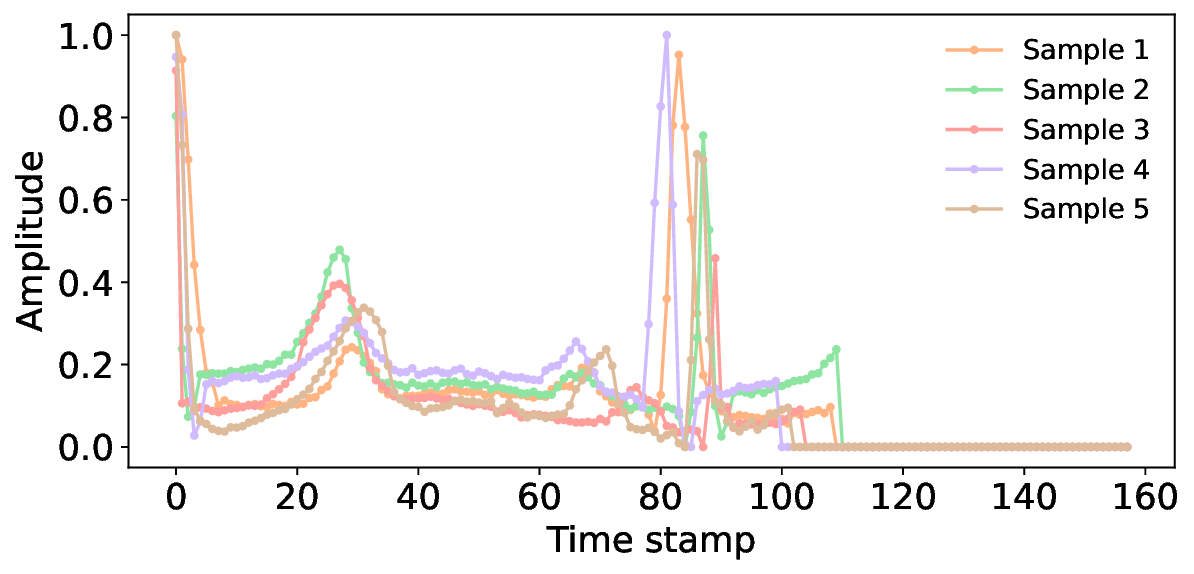}
         \caption{Normal heartbeat}
         \label{fig:norm}
     \end{subfigure}
	 \begin{subfigure}[b]{0.45\textwidth}
         \centering
         \includegraphics[width=\textwidth]{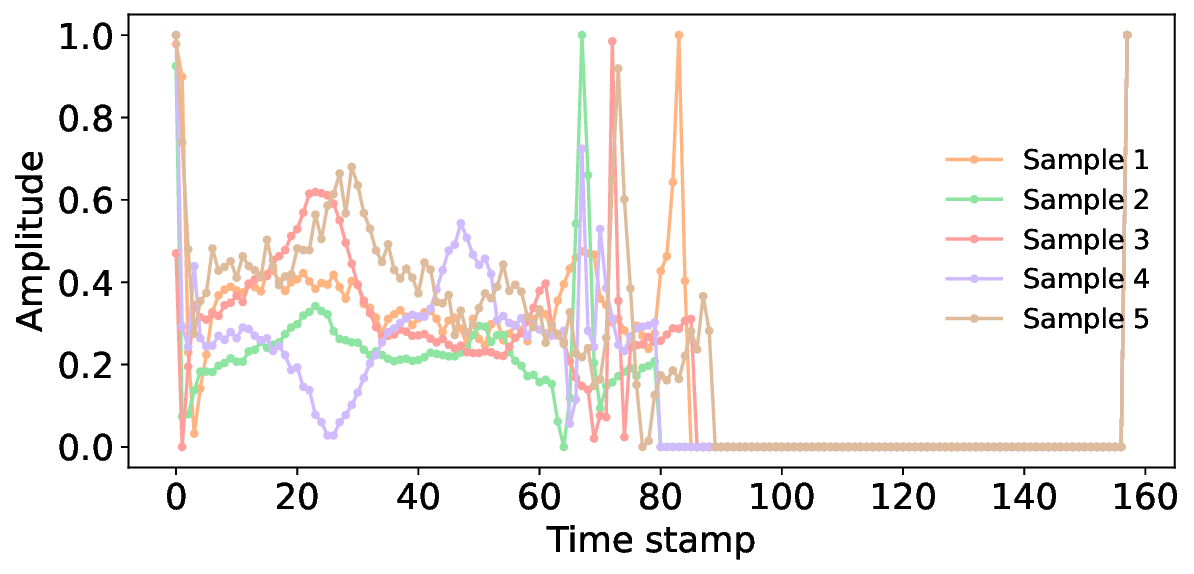}
         \caption{Abnormal heartbeat}
         \label{fig:abnorm}
     \end{subfigure}
     	        \caption{Representation of scanty time series data of the ECG system exhibiting (a) normal and (b) abnormal heartbeats.}
        		\label{fig:ECG_ts}
     \end{figure}

We randomly remove $16\%$ of the data (which accounts for 30 data points) from the original ECG time series signal (which consists of 188 data points),  to obtain scanty ECG signals containing 158 data points. Figure \ref{fig:ECG_ts} presents representative samples of such scanty normal and abnormal ECG signals. Unlike the previous sections, which typically display a single scanty representative signal in each class (one in periodic and the other in chaotic), we present 5 instances of such scanty normal and abnormal ECG signals to showcase the variable dynamics of the signals in both cases. A physician might be able to distinguish the normal and abnormal samples precisely, and they might go a step further by characterizing the abnormal samples presented in Fig. \ref{fig:abnorm} with their respective cardiac diseases. But to a physicist or data scientist, they might seem murky. However, we help the readers to look for a few things that might visually help us to distinguish these two classes (normal and abnormal). Abnormalities in ECG waveforms are characterized by the absence of P waves, inverted T waves, ST depression, and irregular rhythms \cite{ashley2004cardiology,atwood2015ecg}, and these signs are evidently visible in the 5 abnormal samples presented in Fig. \ref{fig:abnorm}.

However, our job is to find whether our proposed methodology picks these features and classifies them into the respective classes. To do this, we stick to the same methodology to compute sublevel set persistence for 4000 ECG time series samples and extract the topological features, after which they are vectorized into 5 input features (following the scheme mentioned in Sec. \ref{vectorization}). We then compile the input features into a master set with labels `0' if it is normal and `1' if it is abnormal.

\begin{figure}
\begin{minipage}{.5\linewidth}
\centering
\subfloat[Confusion matrix for LR]{\label{fig:cnf_LR_ecg}\includegraphics[scale=.22]{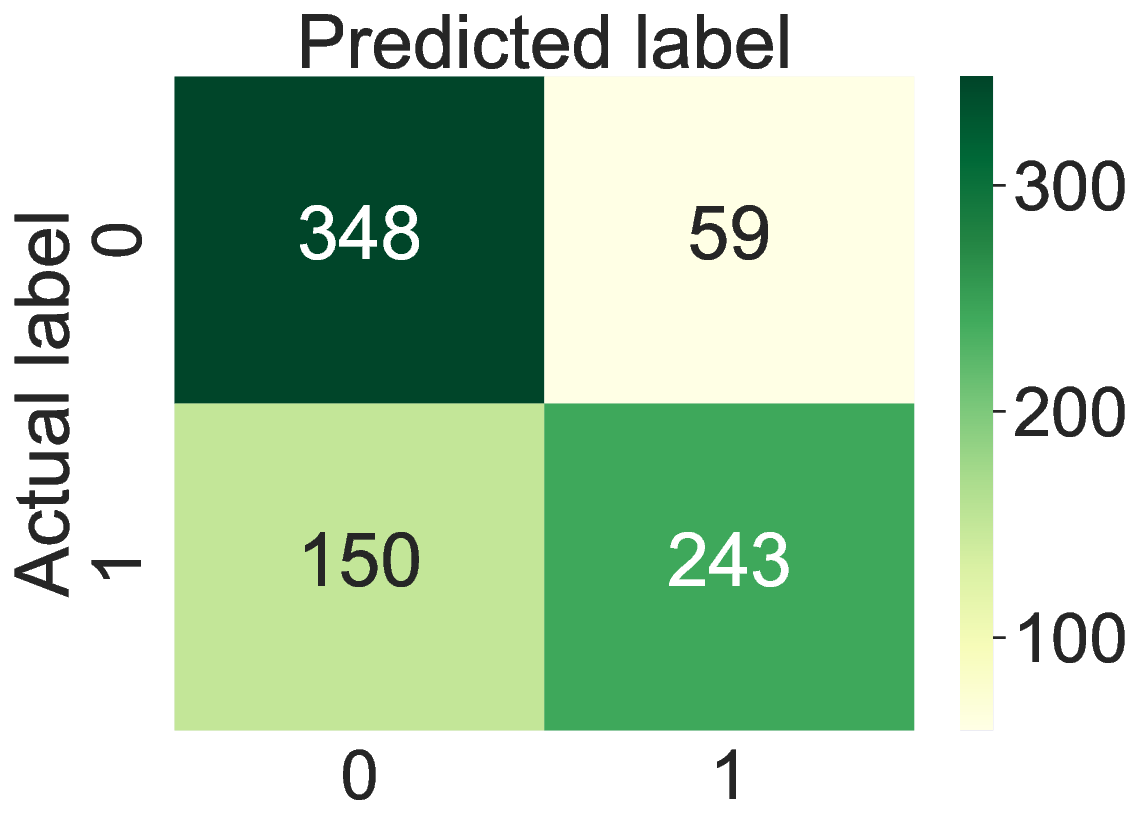}}
\end{minipage}%
\begin{minipage}{.5\linewidth}
\centering
\subfloat[Confusion matrix for SVM]{\label{fig:cnf_SVM_ecg}\includegraphics[scale=.22]{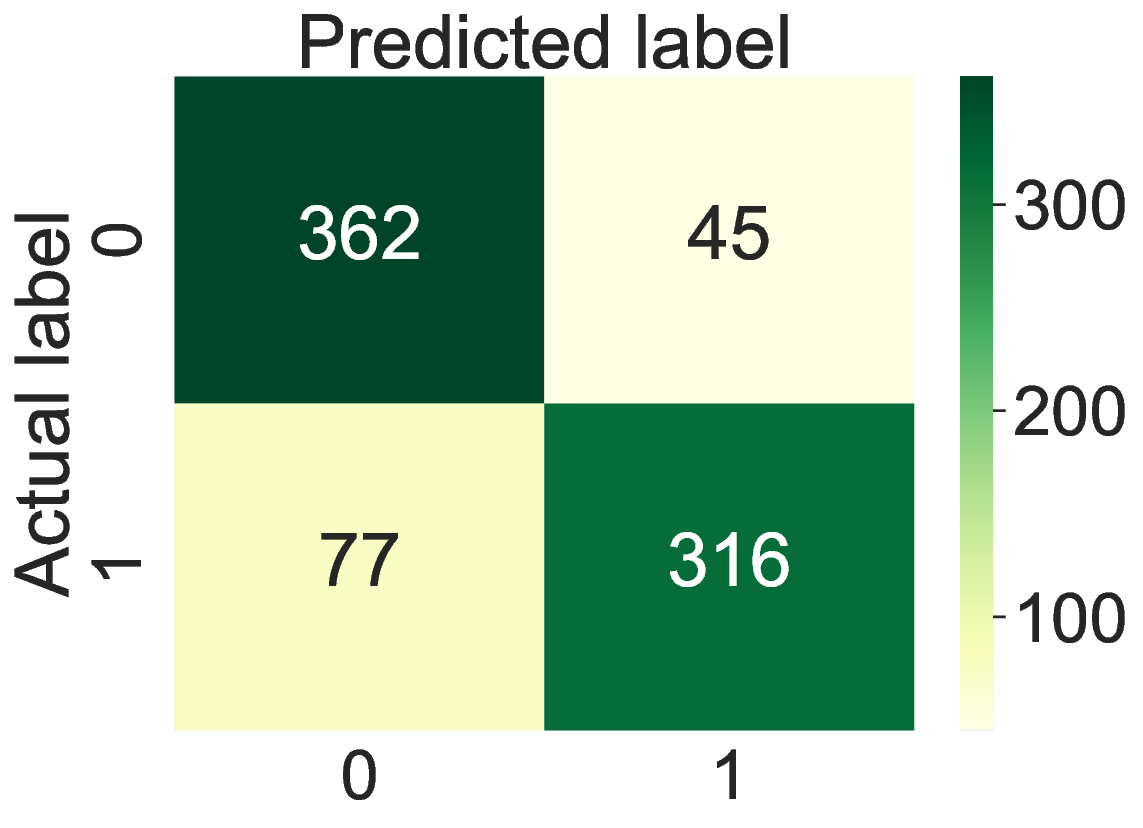}}
\end{minipage}\par\medskip
\centering
\subfloat[Confusion matrix for kNN]{\label{cnf_kNN_ecg}\includegraphics[scale=.22]{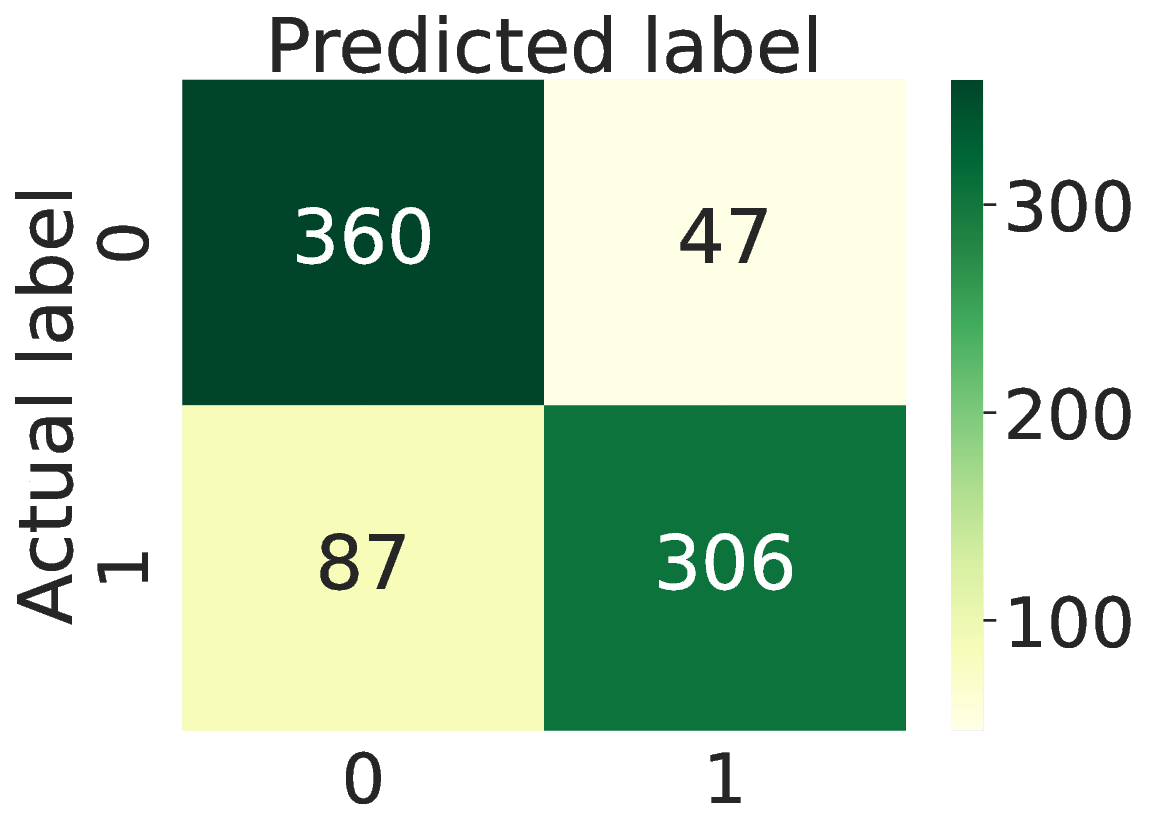}}
\caption{This figure compares the confusion matrix of LR, SVM, and kNN for the ECG data obtained by evaluating the same test set data: We have a test size that accounts for 800 data samples. LR, SVM, and kNN correctly predicted 591, 678, and 666  samples and wrongly predicted 209, 122, and 134 samples, respectively. We see that the SVM and kNN relatively have higher correct predictions than the LR.}
\label{fig:cnf_mat_ECG}
\end{figure}

\begin{figure}
\centering
\includegraphics[width=0.42\textwidth]{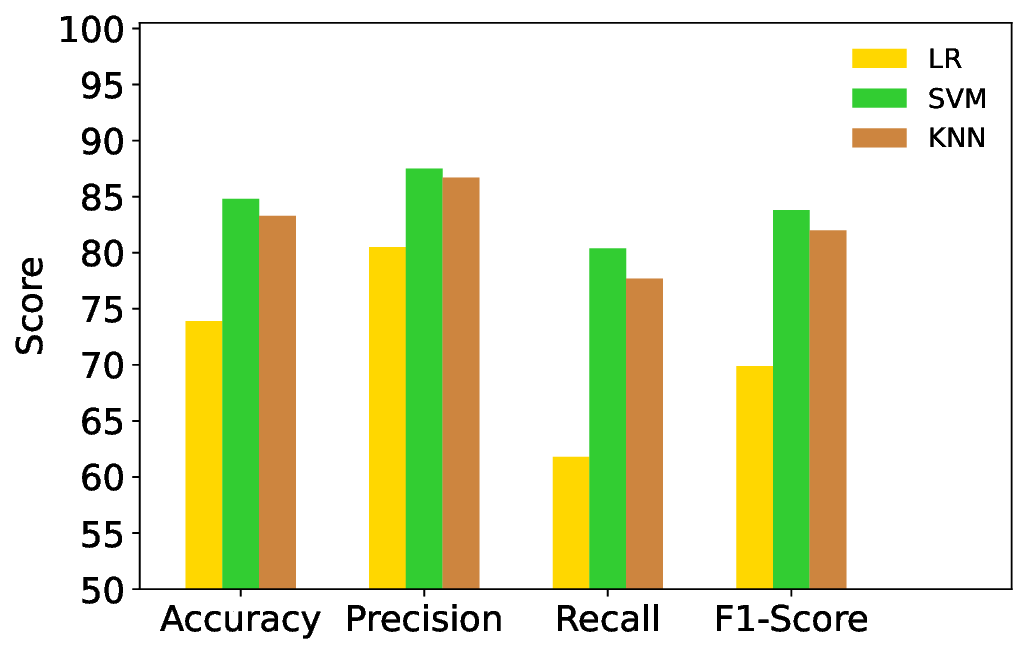}
    \caption{We present the evaluation metric scores of LR, SVM, and kNN for ECG data. They all obtained good scores, showcasing their classification abilities. However, SVM and kNN have outperformed LR in terms of their scores.}
    \label{fig:metrics_ECG}
\end{figure}

Finally, we split the dataset with $80\%$ as the training set and $20\%$ as the test set, and we allow the three ML models to train and find the threshold for classification. The $20\%$ test set data comprises 800 samples. The same set of 800 test samples is exposed to all three classifiers so that they are judged on the same scale. We present the classification results using the confusion matrix for all three classifiers obtained using the test data in Fig. \ref{fig:cnf_mat_ECG}. The summaries of the matrices are: LR made 591 correct predictions, SVM made 678 correct predictions, and kNN made 666 correct predictions. However, LR, SVM, and kNN made 209, 122, and 134 wrong predictions, respectively. We also see their classification abilities translated into the evaluation metrics that are presented in Fig. \ref{fig:metrics_ECG}. The metrics show that both SVM and kNN have outperformed LR. Additionally, we tested the performance of the classifiers over 30 independent random seeds, and the average accuracy obtained for LR is $73.57\% \pm 1.30\%$, SVM is $84.05\% \pm 1.10\%$, and for kNN is $98.47\% \pm 1.43\%$. We would like to acknowledge that the obtained metric scores are not excellent, but we appreciate that our approach demonstrates good classification abilities for real-world data that is laden with missing points. We would also like to admit that the analysis was done on scanty ECG signals, which also plays a factor in the deterioration of the performance metrics

In Table \ref{tabl:hyp_par_ecg}, we present the hyperparameter details for all three classifiers. We then present the 5-fold cross-validation scores obtained using the tuned hyperparameters in Table \ref{tab:table_acc_ecg}. The table presents the accuracy score obtained during each trial along with the average accuracy score of all classifiers. We obtained a consistent score across all 5 trials, suggesting that the chosen hyperparameters generalize well

\begin{table*}
\centering
\begin{tabular}{|ll|ll|ll|}
\hline
\multicolumn{2}{|l|}{\textbf{LR}}                                                  & \multicolumn{2}{l|}{\textbf{SVM}}                                       & \multicolumn{2}{l|}{\textbf{kNN}}                                    \\ \hline
\multicolumn{1}{|l|}{\textbf{Hyper-parameter(s)}}          & \textbf{Value(s)}     & \multicolumn{1}{l|}{\textbf{Hyper-parameter(s)}} & \textbf{Value(s)}    & \multicolumn{1}{l|}{\textbf{Hyper-parameter(s)}} & \textbf{Value(s)} \\ \hline
\multicolumn{1}{|l|}{\multirow{3}{*}{Maximum   iteration}} & \multirow{3}{*}{200} & \multicolumn{1}{l|}{Kernel   function}           & rbf                  & \multicolumn{1}{l|}{No. of   neighbors (k)}      & 5                 \\ \cline{3-6} 
\multicolumn{1}{|l|}{}                                     &                       & \multicolumn{1}{l|}{\multirow{2}{*}{C}}          & \multirow{2}{*}{100} & \multicolumn{1}{l|}{Weights}                     & distance          \\ \cline{5-6} 
\multicolumn{1}{|l|}{}                                     &                       & \multicolumn{1}{l|}{}                            &                      & \multicolumn{1}{l|}{Distance   metric}           & Manhattan         \\ \hline
\end{tabular}
\caption{Table representing hyperparameter details for LR, SVM, and kNN for ECG data.}
\label{tabl:hyp_par_ecg}
\end{table*}

\begin{table}
\centering
\begin{tabular}{|l|lll|}
\hline
\multirow{2}{*}{\textbf{Split (K=5)}} & \multicolumn{3}{l|}{\textbf{Accuracy $\%$}}                                         \\ \cline{2-4} 
                                      & \multicolumn{1}{l|}{\textbf{LR}} & \multicolumn{1}{l|}{\textbf{SVM}}  & \textbf{kNN} \\ \hline
\textbf{Split 1}                      & \multicolumn{1}{l|}{73}         & \multicolumn{1}{l|}{82.8}            & 85.5         \\ \hline
\textbf{Split 2}                      & \multicolumn{1}{l|}{76.1}         & \multicolumn{1}{l|}{86.3}           & 83.9          \\ \hline
\textbf{Split 3}                      & \multicolumn{1}{l|}{73.9}         & \multicolumn{1}{l|}{84.1}          & 83.8          \\ \hline
\textbf{Split 4}                      & \multicolumn{1}{l|}{70.5}        & \multicolumn{1}{l|}{84.4}           & 84.5         \\ \hline
\textbf{Split 5}                      & \multicolumn{1}{l|}{74.7}        & \multicolumn{1}{l|}{84.5}            & 83.4         \\ \hline
\textbf{Average}                      & \multicolumn{1}{l|}{\textbf{73.6}} & \multicolumn{1}{l|}{\textbf{84.4}} & \textbf{84.2}  \\ \hline
\end{tabular}
\caption{We report the accuracy score for each split in 5-fold cross-validation applied for three classifiers: LR, SVM, and kNN for ECG data. The accuracy scores obtained for 5 trials are consistent, and the good average accuracy scores back the hyperparameter choices and indicate that the model generalizes well to unseen data.}
\label{tab:table_acc_ecg}
\end{table}

\section{Conclusion and Discussion:}
\label{sec:Conc}
This study presents a novel approach to detect the dynamic state, viz., identifying the periodic and chaotic phases in dynamical systems from scanty time series or experimental time series laden with missing elements. We devise this methodology utilizing tools from topological data analysis (TDA) and machine learning (ML). Scanty time series or experimental time series with missing elements pose challenges in reconstructing the data, which gave us the motivation to use data-based analysis leveraging a tool from TDA called 0-D Sublevel set persistence. The 0-D sublevel set specifically tracks the emergence of minima and maxima from a real-valued function and maps them as topological features via persistence diagrams (PDs). We utilize this concept in dynamical systems' time series data to extract topological insights based on their behavior. Upon investigation, we observed that periodic time series exhibit repeating minima and maxima, where sublevel set analysis of these yields us overlapping, repeating topological features in the persistence diagram (PD); on the other hand, chaotic time series exhibit non-repeating minima and maxima, where sublevel set analysis of these yields scattered, non-repeating topological features in PD. Now, we require a tool that identifies the repeating topological feature pattern of the periodic states and the non-repeating topological feature pattern of the chaotic states and effectively constructs a decision boundary that separates these two dynamical regimes. We fill this void utilizing the prowess of ML, specifically LR, SVM, and kNN binary classifiers. 

In this paper, we consider scanty time series obtained over a range of bifurcation parameters for a specified step size encompassing both periodic and chaotic states. We then compute sublevel set persistence to extract their topological features obtained from PD. The birth–death coordinate pairs extracted from different time series typically vary in number, resulting in variable-length PDs. To enable their use in machine learning pipelines, we transform them into fixed-length vectors (through a process called "vectorization" or "featurization") using Carlsson's four coordinates ($f_1-f_4$) supplemented by our proposed fifth coordinate ($f_5$), which computes the variance of the pairwise Euclidean distance between the points in PD. Now, these 5 coordinates ($f_1-f_5$) will act as input features (as columns) with each bifurcation parameter acting as a sample (stacked in rows). We label the periodic and chaotic state parameters with target labels of `0' and `1', respectively. We use a $80\%-20\%$ train-test split and use the same random seed for all the classifiers. This ensures that the evaluation metrics used are estimated on the same test set. Furthermore, we tune the hyperparameters of these three classifiers and pick the best set of hyperparameter combinations using the K-fold cross-validation technique. We intentionally use three models with different learning algorithms to compare and contrast classifiers. Subsequently, the performance of these classifiers is judged based on the evaluation metric scores, and we leave it to the practitioners to choose the best-suited one based on the metrics presented. We do not stop with the test set; we go a step further by testing our classifiers on unseen data called \emph{new test parameters}. Our trained, optimized, hyper-tuned models are exposed to a newer range of bifurcation or intermediate range parameters to test how they perform in the real world. We compute sublevel set persistence for the \emph{new test parameters}, and the extracted topological features are vectorized and further sent to the trained classifiers for classification. Our trained models will typically act like binary quantifiers because they assign the dynamical state of \emph{new test parameters} using a binary score: 0 depicting a periodic state, and 1 depicting a chaotic state. This exercise is specifically done to showcase the ability of our optimized models to predict the new, unseen cases that were not included in the labeled dataset. In addition to this, these \emph{new test parameters} allow us to compare the real-world performance of our classifiers and the traditional proven tools like the maximal Lyapunov exponent computed using the Rosenstein method ($\lambda_r$).

We test this proposed methodology on three well-studied systems: two-dimensional systems, such as the Duffing system, and three-dimensional systems, including the Lorenz and Rossler systems. We gauge the classification capabilities of the three classifiers using the evaluation metric scores and compare the performance of the three classifiers to find the best-suited one for the task. We obtained high metric scores for all the systems that were subjected to study. In addition, high metric scores were obtained by tuning the hyperparameters of the classifiers, where we tabulate the hyperparameter details for the classifiers in all the systems for better reproducibility of the research work. Consequently, we employ K-fold cross-validation to substantiate the hyperparameters, show consistent scores for all the trials, and generalize well to the validation set, and we tabulate the respective accuracy score for each trial along with its average accuracy scores for all the systems in Table \ref{tab:table_duff_acc},\ref{tab:table_acc_ross},\ref{tab:table_acc_lor}. In many ML-related studies \cite{subasi2019epileptic,hasan2020diabetes,narudin2016evaluation}, the aim is to develop a classification framework and report its performance using standard evaluation metrics. However, in our work, we go a step further by evaluating the robustness of our classifiers, exposing them to \emph{new test parameters} for three systems: Duffing, Lorenz, and Rossler. This extension of work paves the way to two key implications: First, it enables a direct benchmarking of the dynamical transitions identified by our classifiers against those made by the maximum Lyapunov exponent computed using Bennetin's method, which relies on the governing equations, making it accurate. Second, it facilitates comparison of transitions made by our classifiers with a practical alternative -- which is MLE computed using Rossenstein's method ($\lambda_r$) -- which is the only possible method to compute MLE when only the time series data is available. Consequently, to our surprise, all our classifiers were able to clearly demarcate the dynamical state transition. The transitions made by our classifiers for all the systems exactly match the MLEs and show a clear distinction between the periodic and the chaotic phases. We mark the transcending area using red dashed lines to highlight that they all have the same region of transition. Notably, we would like to insist that, in certain cases (for Lorenz and Rossler), our classifiers, which function as binary quantifiers, outperformed the MLE computed using Rossenstein ($\lambda_r$) by precisely identifying the onset of the transition from periodic to chaotic dynamics.

In addition to the three well-studied dynamical systems, we extended our analysis to a real-world physiological dataset -- ECG time series. To our surprise, we were able to showcase that our framework was able to distinguish healthy or normal ECG time series data from abnormal ECG time series data.  Among the classifiers employed, the SVM classifier produced strong performance with an accuracy score of approximately $85\%$. We would also like to acknowledge that the metrics showcased in Fig. \ref{fig:metrics_ECG} may not represent the state-of-the-art benchmarks in the field. However, it is noteworthy to mention that the analyses were done on sparse or scanty ECG time series data, highlighting the novelty and USP of our approach. 

\begin{table*}[]
\centering
\begin{tabular}{l|ll|}
\cline{2-3}
                                  \multicolumn{1}{l|}{}                  & \multicolumn{2}{c|}{\textbf{Average accuracy score}}                                                                                                                                         \\ \hline
\multicolumn{1}{|c|}{\textbf{Systems}} & \multicolumn{1}{c|}{\textbf{\begin{tabular}[c]{@{}c@{}}TDA + ML \\ (Proposed method)\end{tabular}}} & \textbf{\begin{tabular}[c]{@{}c@{}}0-1 Chaos test \\ (Mainstream method)\end{tabular}} \\ \hline
\multicolumn{1}{|c|}{\textbf{Duffing}} & \multicolumn{1}{c|}{97.84\%   ± 1.92\%}                                                             & 68.63\% ± 5.31\%                                                                       \\ \hline
\multicolumn{1}{|c|}{\textbf{Lorenz}}  & \multicolumn{1}{c|}{96.21\%   ± 2.37\%}                                                             & 51.67\% ± 6.22\%                                                                       \\ \hline
\multicolumn{1}{|c|}{\textbf{Rossler}} & \multicolumn{1}{c|}{99\%   ± 1.34\%}                                                                & 67.73\%   ± 6.98\%                                                                     \\ \hline
\multicolumn{1}{|c|}{\textbf{ECG}}     & \multicolumn{1}{c|}{84.05\%   ± 1.10\%}                                                             & 45.61\% ± 1.35\%                                                                       \\ \hline
\end{tabular}
\caption{Comparison of the average classification accuracy between the proposed TDA+ML framework and the mainstream 0-1 chaos test using the same labeled data. The average accuracy score of the best-performing classifier for each system is presented in the table.}
\label{tab:chaostest_comp}
\end{table*}

Although our obtained results in Figs. \ref{fig:duff_combined_result}, \ref{fig:ross_combined_result},\ref{fig:lor_combined_result} were found to clearly outperform the MLE computed using the Rossenstein method, $\lambda_r$ (despite being $\lambda_r$ computed on regularly sampled time series data), it is imperative to also compare our approach with a methodology that does not involve phase space reconstruction and provides a binary interpretation of dynamical behavior. In this regard, the 0-1 chaos test \cite{gottwald2004new,gottwald2009implementation} will ideally serve as a suitable benchmark, as it is widely adopted in the nonlinear dynamics and engineering community. Hence, we used the same labeled dataset and evaluated the performance of the 0-1 chaos test over 30 independent random seeds, and the computed average classification accuracies are reported in Table. \ref{tab:chaostest_comp}. Our proposed method consistently achieved higher accuracy across all the systems. The lower performance of the 0-1 chaos test may be attributed to its limitation to time series length and sampling inaccuracies (due to the presence of missing data points).

We would like to insist that the entire study was done using the scanty time series data by randomly removing a heavy chunk of data from the uniformly sampled time series data, leaving the data spurious. This is done to reflect or mimic the common nuance encountered in real-world scenarios where data availability is often sparse or constrained. However, the authors would like to disclose that the same methodology can be applied to time series data that are complete or uniformly sampled time series, which constitute a comparatively simpler case. The primary objective of this study is to introduce a framework or methodology for dynamic state change detection of systems where limited and scanty time series data is available, where traditional reconstruction techniques are not feasible. The proposed approach relies on prior knowledge of a finite set of labeled periodic and chaotic trajectories to train supervised machine learning models.

The novelty of the work lies in the application of sublevel set analysis as a feature extraction technique, thereby establishing a bridge between TDA and ML for the effective analysis of nonlinear time series data. Secondly, as mentioned earlier, our methodology demonstrates robustness in scenarios involving sparse or scanty time series data, a context where traditional proven methodologies fail. Our statement is substantiated by the results shown in Sec. \ref{results_lorenz} and Sec. \ref{results_ross}.

We hope that this data-driven methodology will be of great help to the experimentalists dealing with scanty time series. This automated approach also helps in alarming real-world industrial, dynamical, and medical systems operators that undergo phase transitions and prevent the system from further deterioration. This framework involving sublevel set analysis will be particularly useful for protein folding problems \cite{mirth2021representations} in the effective classification of energy landscapes. Additionally, one can employ real-world medical time series data, EEG -- specifically in the identification of seizures -- and also for ECG data in the early prediction of anomalies such as atrial fibrillation. In the future, we would like to investigate the role of noise in the time series and analyze the performance of our models. 

\section{Acknowledgement}
RA would like to thank Dr. Sankar P for the valuable discussions and the insights that helped to construct the annotated dataset. 

\section{Funding}
No funding was received for this work. 

\section{Conflict Of Interest}
The authors of this paper would like to disclose that there is no conflict of interest among us.

\section{Data Availability}
The simulated and the publicly available data \cite{kaggle_ecg} used in this work will be provided upon request to the corresponding author. 

\section{Authors' Contribution}
Rishab Antosh B conceptualized and designed the study, developed the methodology, performed the computational analyses and visualizations in Python, and wrote the original manuscript and revised it. Sanjit Das provided supervision and guidance. Nirmal Thyagu N designed the study, supervised the work, and reviewed and revised the manuscript.

\bibliography{Sub-refs}

\end{document}